\documentclass[11pt]{article}
\usepackage{epsf}
\usepackage{amsmath}
\usepackage{epsfig}
\usepackage{times}
\usepackage{amssymb}
\usepackage{amsthm}
\usepackage{setspace}
\usepackage{cite}

\usepackage{algorithmic}  % This package provides an algorithmic environment fo describing algorithms.
\usepackage{algorithm}

\def\w{{\bf w}}

\def\y{{\bf y}}

\def\x{{\bf x}}

\def\x{{\mathbf x}}

\def\w{{\bf w}}

\def\x{{\bf x}}
\def\y{{\bf y}}
\def\z{{\bf z}}

\def\a{{\bf a}}
\def\b{{\bf b}}

\def\h{{\bf h}}

\def\be{\begin{equation}}
\def\ee{\end{equation}}
\def\ba{\left[\begin{array}}
\def\ea{\end{array}\right]}

\def\w{{\bf w}}

\def\x{{\bf x}}
\def\y{{\bf y}}
\def\z{{\bf z}}

\def\a{{\bf a}}
\def\b{{\bf b}}

\def\1{{\bf 1}}

\def\0{{\bf 0}}

\def\Bstr{B_{s}}
\def\cstr{c_{s}}

\def\erfinv{\mbox{erfinv}}
\def\htheta{\hat{\theta}}

\def\tilh{\tilde{\bf h}}

\def\tila{\tilde{\bf a}}

\def\tilb{\tilde{\bf b}}

\def\deltastr{\delta_s}

\def\Sw{S_w}
\def\Bweak{B_{w}}
\def\hw{\bar{\h}}
\def\cweak{c_{w}}

\def\betasec{\beta_{sec}}

\def\Ssec{S_{sec}}
\def\Bsec{B_{sec}}
\def\hsec{\hat{\h}}
\def\csec{c_{sec}}
\def\deltasec{\delta_{sec}}
\def\hthetasec{\hat{\theta}_{sec}}
\def\thetasec{\theta_{sec}}

\def\Swnon{S_w^{+}}
\def\Bweaknon{B_{weak}^+}
\def\hwnon{\bar{\h}^+}
\def\cweaknon{c_{weak}^+}
\def\betawnon{\beta_{w}^{+}}
\def\deltawnon{\delta_{w}^+}
\def\hthetawnon{\hat{\theta}_{w}^+}
\def\thetawnon{\theta_{w}^+}

\newtheorem{theorem}{Theorem}
\newtheorem{corollary}{Corollary}

\newtheorem{lemma}{Lemma}

\begin{document}

\begin{singlespace}

\title {Various thresholds for $\ell_1$-optimization in compressed
sensing
%\footnote{ This work was supported in
%part.}
}
\author{
\textsc{Mihailo Stojnic}
\\
\\
{School of Industrial Engineering}\\
{Purdue University, West Lafayette, IN 47907} \\
{e-mail: {\tt mstojnic@purdue.edu}} }
\date{July 2009}
\maketitle

\centerline{{\bf Abstract}} \vspace*{0.1in}

Recently, \cite{CRT,DonohoPol} theoretically analyzed the success of
a polynomial $\ell_1$-optimization algorithm in solving an under-determined system of linear equations.
In a large dimensional and
statistical context \cite{CRT,DonohoPol} proved that if the number of equations (measurements
in the compressed sensing terminology) in the system is proportional
to the length of the unknown vector then there is a sparsity (number
of non-zero elements of the unknown vector) also proportional to the
length of the unknown vector such that $\ell_1$-optimization
succeeds in solving the system. In this paper, we provide an alternative performance analysis of
$\ell_1$-optimization and obtain the proportionality constants that in certain cases match or improve on the best
currently known ones from \cite{DonohoPol,DT}.

\vspace*{0.25in} \noindent {\bf Index Terms: Compressed sensing;
$\ell_1$-optimization} .

\end{singlespace}

%%%%%%%%%%%%%%%%%%%%%%%%%%%%%%%%%%%%%%%%%%%%%%%%%%%%%%%%%%%%%%%%%
\section{Introduction}
\label{sec:back}
%%%%%%%%%%%%%%%%%%%%%%%%%%%%%%%%%%%%%%%%%%%%%%%%%%%%%%%%%%%%%%%%%

In last several years the area of compressed sensing has been the
subject of extensive research. The breakthrough results of
\cite{CRT} and \cite{DonohoPol} theoretically demonstrated that in
certain applications (e.g. signal processing in sensor networks) classical sampling at Nyquist rate may not be necessary to
perfectly recover signals. Namely, it turns out that a crucial compressed sensing problem is finding the sparsest
solution of an under-determined system of equations. While this
problem had been known for a long time it is the work of \cite{CRT}
and \cite{DonohoPol} that rigorously proved for the first time that
a sparse enough solution can be recovered by solving a linear program in polynomial time. These
results generated enormous amount of research with possible
applications ranging from high-dimensional geometry, image
reconstruction, single-pixel camera design, decoding of linear
codes, channel estimation in wireless communications, to machine
learning, data-streaming algorithms, DNA micro-arrays,
magneto-encephalography etc. (more on the compressed sensing
problems, their importance, and wide spectrum of different
applications can be found in excellent references
\cite{ECicm,DDTLSKB,CT,JRimaging,BCDH08,CRchannel,VPH,PVMHjournal,WM08,Olgica,RFPrank,MBPSZ08,RS08}).

\begin{figure}[htb]
%%%%%\begin{minipage}[b]{1.0\linewidth}
\centering
\centerline{\epsfig{figure=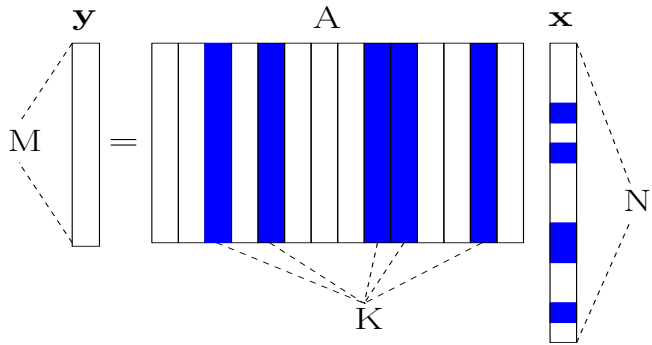,width=9cm,height=4.5cm}}
%%%%%%\end{minipage}
\caption{Model of a linear system; vector $\x$ is $k$-sparse}
\label{fig:model}
\end{figure}

In this paper we are interested in the mathematical background of
certain compressed sensing problems. As is well known, these
problems are very easy to pose and very difficult to solve. Namely,
they are as simple as the following: we would like to find $\x$ such
that
\begin{equation}
A\x=\y \label{eq:system}
\end{equation}
where $A$ is an $m\times n$ ($m<n$) measurement matrix and $\y$ is
an $m\times 1$ measurement vector. Standard compressed sensing
context assumes that $\x$ is an $n\times 1$ unknown $k$-sparse vector (see Figure
\ref{fig:model}; here and in the rest of the paper, under $k$-sparse vector we assume a vector that has at most $k$ nonzero
components). The main topic of this paper will be compressed sensing of the so-called ideally sparse signals (more on the so-called
approximately sparse signals can be found in e.g. \cite{DeVorenoise,
SXHapp,XHapp,StojnicISIT09}). We will mostly throughout the paper assume no special structure on the sparse signal
(more on the very relevant cases of sparse signals with special structures the interested reader can find in \cite{StojnicICASSP09block,StojnicJSTSP09,SPH,FHicassp,EldBol09,EMsub,BCDH08,TGS05,BWDSB05,CH06,CREKD,Temlyakov04,VPH,KDXH09,XKAH09,KXAH09,BerFri09,EKB09,ZeGoAd09,ZeWaSeGoAd08,GaZhMa09,CeInHeBa09,EldRau09,BluDav09,MEldar,NegWai09}).
Also, in the rest of the paper we will assume the
so-called \emph{linear} regime, i.e. we will assume that $k=\beta n$
and that the number of the measurements is $m=\alpha n$ where
$\alpha$ and $\beta$ are absolute constants independent of $n$ (more
on the non-linear regime, i.e. on the regime when $m$ is larger than
linearly proportional to $k$ can be found in e.g.
\cite{CoMu05,GiStTrVe06,GiStTrVe07}).

We generally distinguish two classes of possible algorithms that can be
developed for solving (\ref{eq:system}). The first class of algorithms assumes freedom in designing the measurement matrix $A$.
If one has the freedom to design the measurement matrix $A$ then the results from \cite{FHicassp,Tarokh,MaVe05} demonstrated that the techniques from
coding theory (based on the coding/decoding of Reed-Solomon codes)
can be employed to determine \emph{any} $k$-sparse $\x$ in
(\ref{eq:system}) for any $0<\alpha\leq 1$ and any
$\beta\leq\frac{\alpha}{2}$ in polynomial time. It is relatively easy to show that under the unique recoverability assumption
$\beta$ can not be greater than $\frac{\alpha}{2}$. Therefore, as long as one is concerned with the unique recovery of
$k$-sparse $\x$ in (\ref{eq:system}) in polynomial time the results from \cite{FHicassp,Tarokh,MaVe05} are
optimal. The complexity of algorithms from
\cite{FHicassp,Tarokh,MaVe05} is roughly $O(n^3)$. In a similar fashion one can, instead of using coding/decoding techniques associated with Reed/Solomon codes,
design the measurement matrix and the corresponding recovery algorithm based on the techniques related to the coding/decoding of
Expander codes (see e.g.
\cite{XHexpander,JXHC08,InRu08} and references therein). In that case recovering $\x$ in
(\ref{eq:system}) is significantly faster for large dimensions $n$. Namely, the complexity of the techniques from e.g. \cite{XHexpander,JXHC08,InRu08}
(or their slight modifications) is usually
$O(n)$ which is clearly for large $n$ significantly smaller than $O(n^3)$. However,
the techniques based on coding/decoding of Expander codes usually do not allow for $\beta$ to be as large as
$\frac{\alpha}{2}$.

The main subject of this paper will be the algorithms from the second class. Namely, the second class assumes the algorithms that should be designed without having freedom to design the measurement matrix $A$ in parallel. If one has no freedom in the choice of the matrix $A$ (instead the matrix $A$ is rather given to us) then the recovery
problem (\ref{eq:system}) becomes NP-hard. The following two algorithms (and their different
variations) are then of special interest (and certainly have been the subject of an extensive research in recent years):
\begin{enumerate}
\item \underline{\emph{Orthogonal matching pursuit - OMP}}
\item \underline{\emph{Basis matching pursuit -
$\ell_1$-optimization.}}
\end{enumerate}
Under certain probabilistic assumptions on the elements of the
matrix $A$ it can be shown (see e.g. \cite{JATGomp,JAT,NeVe07,NT08})
that if $\alpha=O(\beta\log(\frac{1}{\beta}))$
OMP (or a slightly modified OMP) can recover $\x$ in (\ref{eq:system})
with complexity of recovery $O(n^2)$. On the other hand the so-called stage-wise
OMP from \cite{DTDSomp} recovers $\x$ in (\ref{eq:system}) with
complexity of recovery $O(n \log n)$.

In this paper we will mostly be interested in the second of the two above mentioned algorithms, i.e. we will be mostly interested in the performance of $\ell_1$-optimization. (Variations of the standard $\ell_1$-optimization from e.g.
\cite{CWBreweighted,SChretien08,SaZh08}) as well as those from \cite{SCY08,FL08,GN03,GN04,GN07,DG08} related to $\ell_q$-optimization, $0<q<1$
are possible as well.) Basic $\ell_1$-optimization algorithm finds $\x$ in
(\ref{eq:system}) by solving the following $\ell_1$-norm minimization problem
\begin{eqnarray}
\mbox{min} & & \|\x\|_{1}\nonumber \\
\mbox{subject to} & & A\x=\y. \label{eq:l1}
\end{eqnarray}
Quite remarkably, in \cite{CT} the authors were able to show that if
$\alpha$ and $n$ are given, the matrix $A$ is given and satisfies a
special property called the restricted isometry property (RIP), then
any unknown vector $\x$ with no more than $k=\beta n$ (where $\beta$
is an absolute constant dependent on $\alpha$ and explicitly
calculated in \cite{CT}) non-zero elements can be recovered by
solving (\ref{eq:l1}). As expected, this assumes that $\y$ was in
fact generated by that $\x$ and given to us. The case when the
available measurements are noisy versions of $\y$ is also of
interest \cite{CT,CRT,HN,W}. Although that case is not of primary interest in the present paper
it is worth mentioning that the recent
popularity of $\ell_1$-optimization in compressed sensing is
significantly due to its robustness with respect to noisy
measurements. (Of course, the main reason for its popularity is its
ability to solve (\ref{eq:system}) for a very wide range of matrices
$A$; more on this universality from a statistical point of view the interested reader can find in \cite{DonTan09Univ}.)

Clearly, having the matrix $A$ satisfy the RIP condition is of
critical importance for previous claim to hold (more on the
importance of the RIP condition can be found in \cite{Crip}). For
several classes of random matrices (e.g., matrices with i.i.d. zero
mean Gaussian, Bernoulli, or even general Sub-gaussian components) the RIP condition is
satisfied with overwhelming probability \cite{CT,Bar,Ver,ALPTJ09}. (Under overwhelming probability we in this paper assume
a probability that is no more than a number exponentially decaying in $n$ away from $1$.) However,
it should be noted that the RIP is only a \emph{sufficient}
condition for $\ell_1$-optimization to produce the solution of
(\ref{eq:system}).

Instead of characterizing the $m\times n$ matrix $A$ through the RIP
condition, in \cite{DonohoUnsigned,DonohoPol} the author associates
certain polytope with the matrix $A$. Namely,
\cite{DonohoUnsigned,DonohoPol} consider polytope obtained by
projecting the regular $n$-dimensional cross-polytope using the
matrix $A$. It turns out that a \emph{necessary and sufficient}
condition for (\ref{eq:l1}) to produce the solution of
(\ref{eq:system}) is that this polytope associated with the matrix
$A$ is $k$-neighborly
\cite{DonohoUnsigned,DonohoPol,DonohoSigned,DT}. Using the results
of \cite{PMM,AS,BorockyHenk,Ruben,VS}, it is further shown in
\cite{DonohoPol}, that if the matrix $A$ is a random $m\times n$
ortho-projector matrix then with overwhelming probability polytope
obtained projecting the standard $n$-dimensional cross-polytope by
$A$ is $k$-neighborly. The precise relation between $m$ and $k$ in
order for this to happen is characterized in
\cite{DonohoPol,DonohoUnsigned} as well.

It should be noted that one usually considers success of
(\ref{eq:l1}) in finding solution of (\ref{eq:system}) for
\emph{any} given $\x$. It is also of interest to consider success of
(\ref{eq:l1}) in finding solution of (\ref{eq:system}) for
\emph{almost any} given $\x$. To make a distinction between these
cases we recall on the following definitions from
\cite{DonohoPol,DT,DTciss}.

Clearly, for any given constant $\alpha\leq 1$ there is a maximum
allowable value of the constant $\beta$ such that (\ref{eq:l1})
finds solution of (\ref{eq:system}) with overwhelming probability
for \emph{any} $\x$. This maximum allowable value of the constant
$\beta$ is called the \emph{strong threshold} (see
\cite{DonohoPol}). We will denote the value of the strong
threshold by $\beta_s$. Similarly, for any given constant
$\alpha\leq 1$ one can define the \emph{sectional threshold} as the
maximum allowable value of the constant $\beta$ such that
(\ref{eq:l1}) finds the solution of (\ref{eq:system}) with overwhelming
probability for \emph{any} $\x$ with a given fixed location of non-zero components (see \cite{DonohoPol}).
In a similar fashion one can then denote the value of the sectional threshold by $\beta_{sec}$. Finally, for any given constant
$\alpha\leq 1$ one can define the \emph{weak threshold} as the
maximum allowable value of the constant $\beta$ such that
(\ref{eq:l1}) finds the solution of (\ref{eq:system}) with overwhelming
probability for \emph{any} $\x$ with a given fixed location of non-zero components and a given fixed combination of its elements signs (see \cite{DonohoPol}).
In a similar fashion one can then denote the value of the weak threshold by $\beta_{w}$. In this paper we determine the values of $\beta_s,\beta_{sec},\beta_w$ for the entire range of $\alpha$, i.e. for $0\leq \alpha\leq 1$, for a specific group of randomly generated matrices $A$.

We organize the rest of the paper in the following way. In Section
\ref{sec:theorems} we introduce two key theorems that will be the heart of our subsequent analysis.
In Section
\ref{sec:unsigned} we determine the values of various thresholds in the case of general sparse signals $\x$ under the assumption that the null-space
of the matrix $A$ is uniformly distributed in the Grassmanian.
Under the same assumption on the statistics of the measurement matrix $A$ in Section \ref{sec:signed} we determine the values of the weak threshold in a special case of the so-called signed vectors $\x$.
Finally, in Section \ref{sec:discuss} we discuss obtained results and
possible directions for future work.

%%%%%%%%%%%%%%%%%%%%%%%%%%%%%%%%%%%%%%%%%%%%%%%%%%%%%%%%%%%%%%%%%
\section{Key theorems}
\label{sec:theorems}
%%%%%%%%%%%%%%%%%%%%%%%%%%%%%%%%%%%%%%%%%%%%%%%%%%%%%%%%%%%%%%%%%

In this section we introduce two useful theorems that will be of key importance in our subsequent
analysis. First we recall on a null-space characterization of
the matrix $A$ that guarantees that the solutions of
(\ref{eq:system}) and (\ref{eq:l1}) coincide. The following theorem from \cite{StojnicICASSP09} provides
this characterization (similar characterizations can be found in
\cite{DH01,FN,LN,Y,XHapp,SPH,DTbern}; furthermore, if instead of $\ell_1$ one, for
example, uses an $\ell_q$-optimization ($0<q<1$) in (\ref{eq:l1}) then
characterizations similar to the ones from
\cite{DH01,FN,LN,Y,XHapp,SPH,DTbern} can be derived as well
\cite{GN03,GN04,GN07}).
\begin{theorem}(Null-space characterization; General $\x$)
Assume that an $m\times n$ measurement matrix $A$ is given. Let $\x$
be a $k$-sparse vector whose non-zero components can be both
positive or negative. Further, assume that $\y=A\x$ and that $\w$ is
an $n\times 1$ vector. Let $K$ be any subset of $\{1,2,\dots,n\}$
such that $|K|=k$ and let $K_i$ denote the $i$-th element of $K$.
Further, let $\bar{K}=\{1,2,\dots,n\}\setminus K$. Let $\1$ be a
$2^k\times k$ sign matrix. Each element of the matrix $\1$ is either
$1$ or $-1$ and there are no two rows that are identical. Let $\1_j$
be the $j$-th row of the matrix $\1$. Then (\ref{eq:l1}) will
produce the solution of (\ref{eq:system}) if
\begin{equation}
(\forall \w\in \textbf{R}^n | A\w=0) \quad \mbox{and} \quad \forall
K,j\quad - \1_j\w_{K}<\sum_{i=1}^{n-k}|\w_{\bar{K}_i}|.
\label{eq:thmeqgen}
\end{equation}
\label{thm:thmgen}
\end{theorem}
\textbf{Remark:} The following simplification of the previous theorem is also well-known. Let $\w\in \textbf{R}^n$ be such that $A\w=0$. Let
$|\w|_{(i)}$ be the $i$-th smallest magnitude of the elements of $\w$. Set $\tilde{\w}=(|\w|_{(1)},|\w|_{(2)},\dots,|\w|_{(n)})^T$.
If $(\forall \w |A\w=0)$
$\sum_{i=n-k+1}^n\tilde{\w}_i\leq \sum_{i=1}^{n-k}\tilde{\w}_i$, where $\tilde{\w}_i$ is the $i$-th element of $\tilde{\w}$, then the solutions of (\ref{eq:system})
and (\ref{eq:l1}) coincide. While we will make use of this formulation in the following section, the
formulation given in Theorem \ref{thm:thmgen} will also turn out to be useful for the analysis that will follow
in the later sections of the paper.

Having matrix $A$ such that
(\ref{eq:thmeqgen}) holds would be enough for solutions of (\ref{eq:l1}) and (\ref{eq:system}) to coincide. If one
assumes that $m$ and $k$ are proportional to $n$ (the case of our
interest in this paper) then the construction of the deterministic
matrices $A$ that would satisfy
(\ref{eq:thmeqgen}) is not an easy task (in fact, one may say that it is one of the most fundamental open problems in the area of theoretical compressed sensing; more on an equally important inverse problem of checking if a given matrix satisfies the condition of Theorem \ref{thm:thmgen} the interested reader can find in \cite{JudNem08,DaEl08}). However, turning to
random matrices significantly simplifies things. As we will see later in the paper, the random matrices $A$ that have the null-space uniformly distributed in the Grassmanian will turn out to be a very convenient choice.
The following phenomenal result from \cite{Gordon88} that relates to such matrices will be the key ingredient in the analysis that will follow.
\begin{theorem}(\cite{Gordon88} Escape through a mesh)
\label{thm:Gordonmesh} Let $S$ be a subset of the unit Euclidean
sphere $S^{n-1}$ in $R^{n}$. Let $Y$ be a random
$(n-m)$-dimensional subspace of $R^{n}$, distributed uniformly in
the Grassmanian with respect to the Haar measure. Let
\begin{equation}
w(S)=E\sup_{\w\in S} (\h^T\w) \label{eq:widthdef}
\end{equation}
where $\h$ is a random column vector in $R^{n}$ with i.i.d. ${\cal
N}(0,1)$ components. Assume that
$w(S)<\left ( \sqrt{m}-\frac{1}{4\sqrt{m}}\right )$. Then
\begin{equation}
P(Y\cap S=0)>1-3.5e^{-\frac{\left (
\sqrt{m}-\frac{1}{4\sqrt{m}}-w(S) \right ) ^2}{18}}.
\label{eq:thmesh}
\end{equation}
\end{theorem}
\textbf{Remark}: Gordon's original constant $3.5$ was substituted by
$2.5$ in \cite{RVmesh}. Both constants are fine for our subsequent
analysis.

%%%%%%%%%%%%%%%%%%%%%%%%%%%%%%%%%%%%%%%%%%%%%%%%%%%%%%%%%%%
\section{Probabilistic analysis of the null-space
characterizations -- general $\x$} \label{sec:unsigned}
%%%%%%%%%%%%%%%%%%%%%%%%%%%%%%%%%%%%%%%%%%%%%%%%%%%%%%%%%%%

In this section we probabilistically analyze validity of the null-space characterization given in Theorem \ref{thm:thmgen}. In the first subsection of this section we will show how one can obtain the values of the strong threshold $\beta_s$ for the entire range $0\leq \alpha\leq 1$ based on such an analysis. In the later two subsections we will generalize the strong threshold analysis and obtain the values of the weak and the sectional threshold.

%%%%%%%%%%%%%%%%%%%%%%%%%%%%%%%%%%%%%%%%%%%%%%%%%%%%%%%%%%%
\subsection{Strong threshold} \label{sec:stronggen}
%%%%%%%%%%%%%%%%%%%%%%%%%%%%%%%%%%%%%%%%%%%%%%%%%%%%%%%%%%%

As masterly noted in \cite{RVmesh} Theorem \ref{thm:Gordonmesh} can
be used to probabilistically analyze (\ref{eq:thmeqgen}) (and as we will see later in the paper, many of its variants).
Namely, let
$S$ in (\ref{eq:widthdef}) be
\begin{equation}
S_s=\{\w\in S^{n-1}|\sum_{i=n-k+1}^n\tilde{\w}_i\leq \sum_{i=1}^{n-k}\tilde{\w}_i\}\label{eq:defS}
\end{equation}
where as earlier the notation $\tilde{\w}$ is used to denote the vector obtained by sorting the absolute values of the elements of $\w$ in non-decreasing order. (Here and later in the paper, we assume that $k$ is chosen such that there is an $0<\alpha\leq 1$ such that the solutions of (\ref{eq:system}) and (\ref{eq:l1}) coincide.) Let $Y$ be an $(n-m)$ dimensional subspace
of $R^{n}$ uniformly distributed in Grassmanian. Furthermore, let
$Y$ be the null-space of $A$. Then as long as
$w(S_s)<\left (\sqrt{m}-\frac{1}{4\sqrt{m}}\right )$, $Y$ will miss $S_s$ (i.e.
(\ref{eq:thmeqgen}) will be satisfied) with probability no smaller than
the one given in (\ref{eq:thmesh}). More precisely, if
$\alpha=\frac{m}{n}$ is a constant (the case of interest in this
paper), $n,m$ are large, and $w(S_s)$ is smaller than but proportional
to $\sqrt{m}$ then $P(Y\cap S_s=0)\longrightarrow 1$. This in turn is
equivalent to having
\begin{equation*}
P(\forall\w \in R^{dn}|A\w=0, \sum_{i=n-k+1}^n\tilde{\w}_i\leq \sum_{i=1}^{n-k}\tilde{\w}_i)\longrightarrow 1
\end{equation*}
which according to Theorem \ref{thm:thmgen} (or more precisely according to the remark after Theorem \ref{thm:thmgen}) means that the solutions
of (\ref{eq:system}) and (\ref{eq:l1}) coincide with probability $1$.
For any given value of $\alpha\in(0,1)$ a threshold value of $\beta$
can be then determined as a maximum $\beta$ such that
$w(S_s)<\left (\sqrt{m}-\frac{1}{4\sqrt{m}}\right )$. That maximum $\beta$ will be exactly the value of the strong threshold $\beta_s$. If one is only concerned with finding a possible value for $\beta_s$ it is easy to note that instead of computing $w(S_s)$ it is sufficient to find its an upper bound. However, as we will see later in the paper, to determine as good values of $\beta_s$ as possible, the upper bound on $w(S_s)$ should be as tight as possible. The main contribution of this work will be a fairly precise estimate of $w(S_s)$.

In the following subsections we present a way to get such an estimate. To simplify the exposition we first set $w(\h,S_s)=\max_{\w\in S_s} (\h^T\w)$. In order to upper-bound $w(S_s)$ we will first in Subsection \ref{sec:strongub} determine an upper bound $\Bstr$ on $w(\h,S_s)$. The expected value with respect to $\h$ of such an upper bound will be an upper bound on $w(S_s)$. In Subsection \ref{sec:strongexp} we will compute an upper bound on that expected value, i.e. we will compute an upper bound on $E (\Bstr)$. That quantity will be an upper bound on $w(S_s)$ since according to the following $E(\Bstr)$ is an upper bound on $w(S_s)$
\begin{equation}
w(S_s)=Ew(\h,S_s)=E(\max_{\w\in S_s} (\h^T\w))\leq E(\Bstr).\label{eq:strgoal}
\end{equation}

%%%%%%%%%%%%%%%%%%%%%%%%%%%%%%%%%%%%%%%%%%%%%%%%%%%%%%%%%%%%%%
\subsubsection{Upper-bounding $w(\h,S_s)$}
\label{sec:strongub}
%%%%%%%%%%%%%%%%%%%%%%%%%%%%%%%%%%%%%%%%%%%%%%%%%%%%%%%%%%%%%%%

From the definition of set $S_s$ given in (\ref{eq:defS}) it easily follows that if $\w$ is in $S_s$ then any vector obtain from $\w$ by changing the signs to any subset of its elements is also in $S_s$. The signs of $\w$ can therefore be chosen so that they match the signs of the corresponding elements in $\h$. We then easily have
\begin{equation}
w(\h,S_s)=\max_{\w\in S_s} (\h^T\w)=\max_{\w\in S_s} \sum_{i=1}^n |\h_i\w_i|=\max_{\w\in S_s} \sum_{i=1}^n |\h_i||\w_i|.\label{eq:workws0}
\end{equation}
Let $|\h|_{(i)}$ be the $i$-th smallest magnitude of elements of $\h$. Set $\tilh=(|\h|_{(1)},|\h|_{(2)},\dots,|\h|_{(n)})^T$. If $\w\in S_s$ then a vector obtained by permuting the elements of $\w$ in any possible way is also in $S_s$. Then (\ref{eq:workws0}) can be rewritten as
\begin{equation}
w(\h,S_s)=\max_{\w\in S_s} \sum_{i=1}^n \tilh_i|\w_i|\label{eq:workws1}
\end{equation}
where $\tilh_i$ is the $i$-th element of vector $\tilh$. We will make use of the following simple lemma.
\begin{lemma}
Let $\hat{\w}$ be the solution of the maximization on the right-hand side of (\ref{eq:workws1}). Then $|\hat{\w}_n|\geq|\hat{\w}_{n-1}|\geq |\hat{\w}_{n-2}|\geq\dots\geq|\hat{\w}_1|$.\label{lm:sortwstr}
\end{lemma}
\begin{proof}
Assume that there is a pair of indexes $n_1,n_2$ such that $n_1< n_2$ and
$|\hat{\w}_{n_1}|>|\hat{\w}_{n_2}|$. However, $|\hat{\w}_{n_1}|\tilh_{n_1}+|\hat{\w}_{n_2}|\tilh_{n_2}<|\hat{\w}_{n_2}|\tilh_{n_1}+|\hat{\w}_{n_1}|\tilh_{n_2}$ and
$\hat{\w}$ would not be the optimal solution of the maximization on the right-hand side of (\ref{eq:workws1}).
\end{proof}
Let $\y=(\y_1,\y_2,\dots,\y_n)^T\in R^n$. Then one can simplify (\ref{eq:workws1}) in the following way
\begin{eqnarray}
w(\h,S_s) = \max_{\y\in R^n} & &  \sum_{i=1}^n \tilh_i\y_i\nonumber \\
\mbox{subject to} &  & \y_i\geq 0, 0\leq i\leq n\nonumber \\
& & \sum_{i=n-k+1}^n\y_i\geq \sum_{i=1}^{n-k} \y_i \nonumber \\
& & \sum_{i=1}^n\y_i^2\leq 1.\label{eq:workws2}
\end{eqnarray}
To be in complete agreement with Lemma \ref{lm:sortwstr} one should add the sorting constraints on the elements of $\y$ in the optimization problem above. However, it is an easy exercise (similar to the proof of Lemma \ref{lm:sortwstr}) to show that these constraints would be redundant, i.e. it is easy to show that any solution $\hat{\y}$ to the above optimization problem will automatically satisfy $\hat{\y}_n\geq\hat{\y}_{n-1}\geq\dots\geq\hat{\y}_1$ (of course since we will be interested in upper-bounding $w(\h,S_s)$ one can even argue that in the maximization problem (\ref{eq:workws1}) dropping constraints would certainly provide a value no smaller than the optimal one obtainable if the constraints are included). To determine an upper bound on $w(\h,S_s)$ we will use the method of Lagrange duality. Before deriving the Lagrange dual we slightly modify (\ref{eq:workws2}) in the following way
\begin{eqnarray}
-w(\h,S_s) = \min_{\y\in R^n} & &  -\sum_{i=1}^n \tilh_i\y_i\nonumber \\
\mbox{subject to} &  & \y_i\geq 0, 0\leq i\leq n\nonumber \\
& & \sum_{i=n-k+1}^n\y_i\geq \sum_{i=1}^{n-k} \y_i \nonumber \\
& & \sum_{i=1}^n\y_i^2\leq 1.\label{eq:workws3}
\end{eqnarray}
To further facilitate writing let $\z\in R^n$ be a column vector such that $\z_i=1,1\leq i\leq (n-k)$ and $\z_i=-1,n-k+1\leq i\leq n$.  Further, let $\lambda=(\lambda_1,\lambda_2,\dots,\lambda_n)^T\in R^n$. Following, e.g. \cite{BV}, we can write the dual of the optimization problem (\ref{eq:workws3}) and its optimal value $w_{up}(\h,S_s)$ as
\begin{eqnarray}
-w_{up}(\h,S_s)=\max_{\gamma,\nu,\lambda} \min_{\y} & & -\tilh^T\y+\gamma||\y||_2^2-\gamma+\nu\z^T\y-\lambda^T\y\nonumber \\
\mbox{subject to} & & \nu\geq 0, \gamma\geq 0\nonumber \\
& & \lambda_i\geq 0, 0\leq i\leq n.
\label{eq:strdual2}
\end{eqnarray}
One can then transform the objective function in the following way
\begin{eqnarray}
-w_{up}(\h,S_s)=\max_{\gamma,\nu,\lambda} \min_{\y} & & \|\sqrt{\gamma}\y-\frac{\lambda+\tilh-\nu\z}{2\sqrt{\gamma}}\|_2^2-\gamma -\frac{\|\lambda+\tilh-\nu\z\|_2^2}{4\gamma}\nonumber \\
\mbox{subject to} & & \nu\geq 0, \gamma\geq 0\nonumber \\
& & \lambda_i\geq 0, 0\leq i\leq n.
\label{eq:strdual3}
\end{eqnarray}
After trivially solving the inner minimization in (\ref{eq:strdual3}) we obtain
\begin{eqnarray}
w_{up}(\h,S_s)=\min_{\gamma,\nu,\lambda} & & \gamma +\frac{\|\lambda+\tilh-\nu\z\|_2^2}{4\gamma}\nonumber \\
\mbox{subject to} & & \nu\geq 0, \gamma\geq 0\nonumber \\
& & \lambda_i\geq 0, 0\leq i\leq n.
\label{eq:strdual4}
\end{eqnarray}
Minimization over $\gamma$ is straightforward and one easily obtains that $\gamma=\frac{\|\lambda+\tilh-\nu\z\|_2}{2}$ is optimal. Plugging this value of $\gamma$ back in the objective function of the optimization problem (\ref{eq:strdual4}) one obtains
\begin{eqnarray}
w_{up}(\h,S_s)=\min_{\nu,\lambda} & & \|\lambda+\tilh-\nu\z\|_2\nonumber \\
\mbox{subject to} & & \nu\geq 0\nonumber \\
& & \lambda_i\geq 0, 0\leq i\leq n.
\label{eq:strdual5}
\end{eqnarray}
By duality, $-w_{up}(\h,S_s)\leq -w(\h,S_s)$ which easily implies $w(\h,S_s)\leq w_{up}(\h,S_s)$. Therefore $w_{up}(\h,S_s)$ is an upper bound on $w(\h,S_s)$. (In fact one can easily show that the strong duality holds and that $w(\h,S_s)=w_{up}(\h,S_s)$; however, as explained earlier, for our analysis showing that $w_{up}(\h,S_s)$ is an upper bound on $w(\h,S_s)$ is sufficient.) Along the same lines, one can easily spot that any feasible values $\nu$ and $\lambda$ in (\ref{eq:strdual5}) will provide a valid upper bound on $w_{up}(\h,S_s)$ and hence a valid upper bound on $w(\h,S_s)$. In what follows we will in fact determine the optimal values for $\nu$ and $\lambda$. However, since it is not necessary for our analysis we will not put too much effort into proving that these values are optimal. As we have stated earlier, for our analysis it will be enough to show that the values for $\nu$ and $\lambda$ that we will obtain are feasible in (\ref{eq:strdual5}).

To facilitate the exposition in what follows instead of dealing with the objective function given in (\ref{eq:strdual5}) we will be dealing with its squared value. Hence, we set $f(\h,\nu,\lambda)=\|\lambda+\tilh-\nu\z\|_2^2$. Now, let $\lambda=(\lambda_1,\lambda_2,\dots,\lambda_c,0,0,\dots,0)^T, \lambda_1\geq\lambda_2\geq\dots\geq\lambda_c\geq 0$ where $c\leq (n-k)$ is a crucial parameter that will be determined later. The optimization over $\nu$ in (\ref{eq:strdual5}) is then seemingly straightforward. Setting the derivative of $f(\h,\nu,\lambda)$ with respect to $\nu$ to zero we have
\begin{eqnarray}
& & \frac{d\|\lambda+\tilh-\nu\z\|_2^2}{d\nu}  =  0\nonumber \\
& \Leftrightarrow &-2(\lambda+\tilh)^T\z+2\|\z\|_2^2\nu=0\nonumber \\
& \Leftrightarrow & \nu=\frac{(\lambda+\tilh)^T\z}{\|\z\|_2^2}.\label{eq:nu}
\end{eqnarray}
If $(\lambda+\tilh)^T\z\geq 0$ then $\nu=\frac{(\lambda+\tilh)^T\z}{\|\z\|_2^2}$ is indeed the optimal in (\ref{eq:strdual5}). For the time being let us assume that $\lambda,\h,c$ are such that $\nu=\frac{(\lambda+\tilh)^T\z}{\|\z\|_2^2}\geq 0$. For $\nu=\frac{(\lambda+\tilh)^T\z}{\|\z\|_2^2}$ we have
\begin{equation}
f(\h,\frac{(\lambda+\tilh)^T\z}{\|\z\|_2^2},\lambda)=\|(\lambda+\tilh)^T(I-\frac{\z\z^T}{\z^T\z})\|_2^2=(\lambda+\tilh)^T(I-\frac{\z\z^T}{\z^T\z})(\lambda+\tilh).
\label{eq:strf1}
\end{equation}
Simplifying (\ref{eq:strf1}) further we obtain
\begin{equation}
f(\h,\frac{(\lambda+\tilh)^T\z}{\|\z\|_2^2},\lambda)=\sum_{i=1}^n\tilh_i^2+2\sum_{i=1}^c\lambda_i\tilh_i+\sum_{i=1}^c\lambda_i^2-\frac{(\tilh^T\z)^2}{n}
-\frac{(\sum_{i=1}^c\lambda_i)^2}{n}-\frac{2(\sum_{i=1}^c\lambda_i)(\tilh^T\z)}{n}.
\label{eq:strf2}
\end{equation}
To determine good values for $\lambda$ we proceed by setting the derivatives of $f(\h,\frac{(\lambda+\tilh)^T\z}{\|\z\|_2^2},\lambda)$ with respect to $\lambda_i,1\leq i\leq c$ to zero
\begin{equation}
\frac{df(\h,\frac{(\lambda+\tilh)^T\z}{\|\z\|_2^2},\lambda)}{d\lambda_i}=2\lambda_i+2\tilh_i-2\frac{(\sum_{i=1}^c\lambda_i)}{n}-2\frac{(\tilh^T\z)}{n}=0.
\label{eq:strdf1}
\end{equation}
Summing the above derivatives over $i$ and equalling with zero we obtain
\begin{equation}
\sum_{i=1}^c\frac{df(\h,\frac{(\lambda+\tilh)^T\z}{\|\z\|_2^2},\lambda)}{d\lambda_i}=2(\sum_{i=1}^c\lambda_i+\sum_{i=1}^c\tilh_i
-c\frac{(\sum_{i=1}^c\lambda_i)}{n}-c\frac{(\tilh^T\z)}{n})=0.
\label{eq:strdf2}
\end{equation}
From (\ref{eq:strdf2}) one then easily finds
\begin{equation}
\sum_{i=1}^c\lambda_i=\frac{c(\tilh^T\z)}{n-c}-\frac{n\sum_{i=1}^c\tilh_i}{n-c}.\label{eq:sumlam1}
\end{equation}
Plugging the value for $\sum_{i=1}^c\lambda_i$ obtained in (\ref{eq:sumlam1}) in (\ref{eq:strdf1}) we have
\begin{equation*}
\lambda_i=\frac{(\tilh^T\z)}{n}-\tilh_i+\frac{(\sum_{i=1}^c\lambda_i)}{n}=\frac{(\tilh^T\z)}{n}-\tilh_i+
\frac{c(\tilh^T\z)}{n(n-c)}-\frac{\sum_{i=1}^c\tilh_i}{n-c}
\end{equation*}
and finally
\begin{eqnarray}
\lambda_i & = & \frac{(\tilh^T\z)-\sum_{i=1}^c\tilh_i}{n-c}-\tilh_i,1\leq i\leq c\nonumber \\
\lambda_i & = & 0,c+1\leq i\leq n.\label{eq:sumlam2}
\end{eqnarray}
Combining (\ref{eq:nu}) and (\ref{eq:sumlam1}) we have
\begin{equation}
\nu=\frac{(\lambda+\tilh)^T\z}{\|\z\|_2^2}=\frac{\tilh^T\z+\sum_{i=1}^c\lambda_i}{n}=
\frac{\tilh^T\z+\frac{c(\tilh^T\z)}{n-c}-\frac{n\sum_{i=1}^c\tilh_i}{n-c}}{n}=\frac{(\tilh^T\z)-\sum_{i=1}^c\tilh_i}{n-c}.\label{eq:finalnu}
\end{equation}
From (\ref{eq:sumlam2}) we then have as expected
\begin{equation}
\nu=\lambda_i+\tilh_i, 1\leq i\leq c.\label{eq:nulam}
\end{equation}
As long as we can find a $c$ such that $\lambda_i,1\leq i\leq c$ given in (\ref{eq:sumlam2}) are non-negative $\nu$ will be non-negative as well and $\nu$ and $\lambda$ will therefore be feasible in (\ref{eq:strdual5}). This in turn implies
\begin{equation}
w(\h,S_s)\leq \sqrt{f(\h,\nu,\lambda)}\label{eq:strupbound0}
\end{equation}
where $f(\h,\nu,\lambda)$ is computed for the values of $\lambda$ and $\nu$ given in (\ref{eq:sumlam2}) and (\ref{eq:nulam}), respectively. (In fact determining the largest $c$ such that $\lambda_i,1\leq i\leq c$ given in (\ref{eq:sumlam2}) are non-negative will insure that
$\sqrt{f(\h,\nu,\lambda)}=w(\h,S_s)$; however, as already stated earlier, this fact is not of any special importance for our analysis).

Let us now assume that $c$ is fixed such that $\lambda$ and $\nu$ are as given in (\ref{eq:sumlam2}) and (\ref{eq:nulam}). Then combining (\ref{eq:strf2}), (\ref{eq:sumlam1}), and (\ref{eq:nulam}) we have
\begin{equation}
f(\h,\frac{(\lambda+\tilh)^T\z}{\|\z\|_2^2},\lambda)=\sum_{i=1}^n\tilh_i^2+2\nu\sum_{i=1}^{c}\tilh_i
-2\sum_{i=1}^{c}\tilh_i^2+c\nu^2-2\nu\sum_{i=1}^{c}\tilh_i+\sum_{i=1}^{c}\tilh_i^2
-\frac{(\sum_{i=1}^{c}\lambda_i+\tilh^T\z)^2}{n}.\label{eq:strupbound1}
\end{equation}
Combining (\ref{eq:sumlam1}) and (\ref{eq:finalnu}) we obtain
\begin{equation}
(\sum_{i=1}^{c}\lambda_i+\tilh^T\z)=n\nu.\label{eq:strupbound2}
\end{equation}
Further, combining (\ref{eq:strupbound1}) and (\ref{eq:strupbound2}) we find
\begin{eqnarray}
f(\h,\frac{(\lambda+\tilh)^T\z}{\|\z\|_2^2},\lambda) & = & \sum_{i=1}^n\tilh_i^2+c\nu^2-\sum_{i=1}^{c}\tilh_i^2-\frac{(n\nu)^2}{n}\nonumber \\
& = & \sum_{i=1}^n\tilh_i^2+(c-n)\nu^2-\sum_{i=1}^{c}\tilh_i^2\nonumber \\
& = & \sum_{i=1}^n\tilh_i^2-\sum_{i=1}^{c}\tilh_i^2-\frac{((\tilh^T\z)-\sum_{i=1}^{c}\tilh_i)^2}{n-c}.\label{eq:strupbound3}
\end{eqnarray}
Finally, combining (\ref{eq:strupbound0}) and (\ref{eq:strupbound3}) we have
\begin{equation}
w(\h,S_s)\leq\sqrt{\sum_{i=1}^n\tilh_i^2-\sum_{i=1}^{c}\tilh_i^2-\frac{((\tilh^T\z)-\sum_{i=1}^{c}\tilh_i)^2}{n-c}}
=\sqrt{\sum_{i=c+1}^n\tilh_i^2-\frac{((\tilh^T\z)-\sum_{i=1}^{c}\tilh_i)^2}{n-c
}}.\label{eq:strfinalub}
\end{equation}
Clearly, as long as $(\tilh^T\z)\geq 0$ there will be a $c\leq n-k$ (it is possible that $c=0$) such that quantity on the most right hand side of
(\ref{eq:strfinalub}) is an upper bound on $w(\h,S_s)$.

To facilitate the exposition in the following subsection we will make the upper bound given in (\ref{eq:strfinalub}) slightly more pessimistic in the following lemma.
\begin{lemma}
Let $\h\in R^n$ be a vector with i.i.d. zero-mean unit variance gaussian components. Further let $|\h|_{(i)},\1\leq i\leq n$, be the $i$-th smallest magnitude of elements of $\h$. Set $\tilh=(|\h|_{(1)},|\h|_{(2)},\dots,|\h|_{(n)})^T$ and $w(\h,S_s)=\max_{\w\in S_s} (\h^T\w)$ where $S_s$ is as defined in (\ref{eq:defS}). Let $\z\in R^n$ be a column vector such that $\z_i=1,1\leq i\leq (n-k)$ and $\z_i=-1,n-k+1\leq i\leq n$. Then
\begin{equation}
w(\h,S_s)\leq \Bstr \label{eq:stronglemmaub}
\end{equation}
where
\begin{equation}
\Bstr=
\begin{cases}
\sqrt{\sum_{i=1}^n\tilh_i^2} & \mbox{if} \quad \zeta_s(\h,\cstr)\leq 0 \\
\sqrt{\sum_{i=\cstr+1}^n\tilh_i^2-\frac{((\tilh^T\z)-\sum_{i=1}^{\cstr}\tilh_i)^2}{n-\cstr}} & \mbox{if} \quad \zeta_s(\h,\cstr)>0
\end{cases},
\end{equation}
$\zeta_s(\h,c)=\frac{(\tilh^T\z)-\sum_{i=1}^c\tilh_i}{n-c}-\tilh_c$
and $\cstr=\deltastr n$ is a $c\leq n-k$ such that
\begin{equation}
\frac{(1-\epsilon)E((\tilh^T\z)-\sum_{i=1}^c\tilh_i)}{n-c}-F_a^{-1}\left (\frac{(1+\epsilon)c}{n}\right )= 0.
\end{equation}
$F_a^{-1}(\cdot)$ is the inverse cdf of the random variable $|X|$ where $X$ is zero-mean, unit variance gaussian random variable. $\epsilon>0$ is an arbitrarily small constant independent of $n$.\label{lm:Bstr}
\end{lemma}
\begin{proof}
Follows from the previous analysis and (\ref{eq:strfinalub}).
\end{proof}

%%%%%%%%%%%%%%%%%%%%%%%%%%%%%%%%%%%%%%%%%%%%%
\subsubsection{Computing  an upper bound on $E(\Bstr)$} \label{sec:strongexp}
%%%%%%%%%%%%%%%%%%%%%%%%%%%%%%%%%%%%%%%%%%%%%

In this subsection we will compute an upper bound on $E(\Bstr)$. As a first step we determine a lower bound on $P(\zeta_s(\h,\cstr)>0)$. We start by a sequence of obvious inequalities
\begin{multline}
P(\zeta_s(\h,\cstr)>0)\geq P\left (\zeta_s(\h,\cstr)\geq \frac{(1-\epsilon)E((\tilh^T\z)-\sum_{i=1}^{\cstr}\tilh_i)}{n-\cstr}-F_a^{-1}\left (\frac{(1+\epsilon)\cstr}{n}\right )\right )\\
\geq P\left ( \frac{((\tilh^T\z)-\sum_{i=1}^{\cstr}\tilh_i)}{n-\cstr}\geq\frac{(1-\epsilon)E((\tilh^T\z)-\sum_{i=1}^{\cstr}\tilh_i)}{n-\cstr}\hspace{.1in} \mbox{and}
\hspace{.1in} F_a^{-1}\left (\frac{(1+\epsilon)\cstr}{n}\right ) \geq \tilh_{\cstr} \right )\\
\geq 1-P\left ( \frac{((\tilh^T\z)-\sum_{i=1}^{\cstr}\tilh_i)}{n-\cstr}<\frac{(1-\epsilon)E((\tilh^T\z)-\sum_{i=1}^{\cstr}\tilh_i)}{n-\cstr}\right )
-P\left (F_a^{-1}\left (\frac{(1+\epsilon)\cstr}{n}\right ) < \tilh_{\cstr} \right )\label{eq:strexp1}
\end{multline}
The rest of the analysis assumes that $n$ is large so that $\deltastr$ can be assumed to be real (of course, $\deltastr$ is a proportionality constant independent of $n$).
Using the results from \cite{BarvinokOS} we obtain
\begin{eqnarray}
P\left (F_a^{-1}\left (\frac{(1+\epsilon)\cstr}{n}\right ) < \tilh_{\cstr} \right ) & \leq & \exp\left \{-\frac{n}{2\frac{(1+\epsilon)\cstr}{n}}
\left ( \frac{\cstr}{n}- \frac{(1+\epsilon)\cstr}{n} \right )^2 \right \}\nonumber \\
& \leq & \exp\left \{-\frac{n\epsilon^2\deltastr}{2(1+\epsilon)} \right \}.\label{eq:strexp2}
\end{eqnarray}
We will also need the following brilliant result from \cite{CIS76}. Let $\xi(\cdot):R^n\longrightarrow R$ be a Lipschitz function such that $|\xi(\a)-\xi(\b)|\leq\sigma\|\a-\b\|_2$. Let $\a$ be a vector comprised of i.i.d. zero-mean, unit variance Gaussian random variables. Then
\begin{equation}
P((1-\epsilon)E\xi(\a)\geq \xi(\a))\leq \exp \left \{  -\frac{(\epsilon E\xi(\a))^2}{2\sigma^2} \right \}.\label{eq:lipsch}
\end{equation}
Let $\xi(\h)=(\tilh^T\z)-\sum_{i=1}^{\cstr}\tilh_i$. The following lemma estimates $\sigma$ (for simplicity we assume $\cstr=0$; the proof easily extends to the case when $\cstr\neq 0$).

\begin{lemma}
Let $\a,\b\in R^n$. Let $|\a|_{(i)},|\b|_{(i)}$ be the $i$-th smallest magnitudes of $\a,\b$, respectively. Set $\tila=(|\a|_{(1)},|\a|_{(2)},\dots,|\a|_{(n)})$
and $\tilb=(|\b|_{(1)},|\b|_{(2)},\dots,|\b|_{(n)})$. Then
\begin{equation}
|\xi(\a)-\xi(\b)|=|\sum_{i=1}^{n-k}\tila_i-\sum_{n-k+1}^n\tila_i-\sum_{i=1}^{n-k}\tilb_i+\sum_{n-k+1}^n\tilb_i|
\leq \sqrt{n}\sqrt{\sum_{i=1}^n|\a_i-\b_i|^2}=\sqrt{n}\|\a-\b\|_2.\label{eq:Liplemma}
\end{equation}\label{lm:Lipschlemma}
\end{lemma}
\begin{proof}
The following sequence of inequalities/equalities is easy to establish
\begin{eqnarray}
& & |\sum_{i=1}^{n-k}\tila_i-\sum_{i=n-k+1}^n\tila_i-\sum_{i=1}^{n-k}\tilb_i+\sum_{i=n-k+1}^n\tilb_i|  \leq
|\sum_{i=1}^{n-k}(\tila_i-\tilb_i)|+|\sum_{i=n-k+1}^{n-k}(\tila_i-\tilb_i)|\nonumber \\
& \leq & \sum_{i=1}^{n-k}|\tila_i-\tilb_i|+\sum_{i=n-k+1}^{n-k}|\tila_i-\tilb_i|\leq
\sum_{i=1}^{n}|\tila_i-\tilb_i| \leq \sqrt{n}\sqrt{\sum_{i=1}^{n}|\tila_i-\tilb_i|^2}\nonumber \\
& \leq & \sqrt{n}\sqrt{\sum_{i=1}^{n}|\tila_i|^2+\sum_{i=1}^{n}|\tilb_i|^2-2\sum_{i=1}^n\tila_i\tilb_i}
= \sqrt{n}\sqrt{\sum_{i=1}^{n}|\a_i|^2+\sum_{i=1}^{n}|\b_i|^2-2\sum_{i=1}^n\tila_i\tilb_i}\nonumber\\
& \leq & \sqrt{n}\sqrt{\sum_{i=1}^{n}|\a_i|^2+\sum_{i=1}^{n}|\b_i|^2-2\sum_{i=1}^n \a_i \b_i}=
\sqrt{n}\sqrt{\sum_{i=1}^n|\a_i-\b_i|^2}.\label{eq:proofLiplemma1}
\end{eqnarray}
The last inequality follows since the components of $\tila$ and $\tilb$ are positive and sorted in the same non-decreasing order. Connecting beginning and end in (\ref{eq:proofLiplemma1}) establishes (\ref{eq:Liplemma}).
\end{proof}
For $\xi(\h)=(\tilh^T\z)-\sum_{i=1}^{\cstr}\tilh_i$ the previous lemma then gives $\sigma\leq\sqrt{n}$ (in fact $\sigma\leq\sqrt{n-\cstr}$).
If $n$ is large and $\deltastr$ is a constant independent of $n$, one can following \cite{Stigler73} (as we will see later in the paper) show that $E((\tilh^T\z)-\sum_{i=1}^{\cstr}\tilh_i)=\psi_s n$ where $\psi_s$ is independent of $n$ as well ($\psi_s$ is of course dependent on $\beta$ and $\deltastr$).
Hence choosing $\xi(\h)=(\tilh^T\z)-\sum_{i=1}^{\cstr}\tilh_i$ in (\ref{eq:lipsch}) we obtain
\begin{equation}
P\left ( \frac{((\tilh^T\z)-\sum_{i=1}^{\cstr}\tilh_i)}{n-\cstr}<\frac{(1-\epsilon)E((\tilh^T\z)-\sum_{i=1}^{\cstr}\tilh_i)}{n-\cstr}\right )
\leq \exp \left \{  -\frac{(\epsilon \psi_s n)^2}{2n} \right \}= \exp \left \{  -\frac{\epsilon^2 \psi_s^2 n}{2} \right \}.\label{eq:strexp3}
\end{equation}
From (\ref{eq:strexp1}), (\ref{eq:strexp2}), and (\ref{eq:strexp3}) we finally have
\begin{multline}
P(\zeta_s(\h,\cstr)>0)\geq 1-P\left ( \frac{((\tilh^T\z)-\sum_{i=1}^{\cstr}\tilh_i)}{n-\cstr}<\frac{(1-\epsilon)E((\tilh^T\z)-\sum_{i=1}^{\cstr}\tilh_i)}{n-\cstr}\right )\\
-P\left (F_a^{-1}\left (\frac{(1+\epsilon)\cstr}{n}\right ) < \tilh_{\cstr} \right )\\
\geq 1-\exp\left \{-\frac{n\epsilon^2\deltastr}{2(1+\epsilon)} \right \}-\exp \left \{  -\frac{\epsilon^2 \psi_s^2 n}{2} \right \}.\label{eq:strexp4}
\end{multline}

We now return to computing an upper bound on $E(\Bstr)$. By the definition of $\Bstr$ we have
\begin{equation}
E(\Bstr)=\int_{\zeta_s(\h,\cstr)\leq 0}\sqrt{\sum_{i=1}^n\tilh_i^2}p(\h)d\h+
\int_{\zeta_s(\h,\cstr)> 0} \sqrt{\sum_{i=\cstr+1}^n\tilh_i^2-\frac{((\tilh^T\z)-\sum_{i=1}^{\cstr}\tilh_i)^2}{n-\cstr}} p(\h)d\h\label{eq:strexp5}
\end{equation}
where $p(\h)$ is the joint pdf of the i.i.d. zero-mean unit variance gaussian components of vector $\h$. Since the functions $\sqrt{\sum_{i=1}^n\tilh_i^2}$ and $p(\h)$ are rotationally invariant and since the region $\zeta_s(\h,\cstr)\leq 0$ takes up the same fraction of the surface area of sphere of any radius we have
\begin{equation}
\int_{\zeta_s(\h,\cstr)\leq 0}\sqrt{\sum_{i=1}^n\tilh_i^2}p(\h)d\h=E\sqrt{\sum_{i=1}^n\tilh_i^2}\int_{\zeta_s(\h,\cstr)\leq 0}p(\h)d\h\leq
\sqrt{E\sum_{i=1}^n\tilh_i^2}\int_{\zeta_s(\h,\cstr)\leq 0}p(\h)d\h.\label{eq:strexp6}
\end{equation}
Combining (\ref{eq:strexp4}) and (\ref{eq:strexp6}) we further have
\begin{equation}
\int_{\zeta_s(\h,\cstr)\leq 0}\sqrt{\sum_{i=1}^n\tilh_i^2}p(\h)d\h\leq \sqrt{E\sum_{i=1}^n\tilh_i^2}
\left (\exp\left \{-\frac{n\epsilon^2\deltastr}{2(1+\epsilon)} \right \}+\exp \left \{  -\frac{\epsilon^2 \psi_s^2 n}{2} \right \}\right ).\label{eq:strexp7}
\end{equation}
It also easily follows
\begin{eqnarray}
& & \hspace{-.4in}\int_{\zeta_s(\h,\cstr)> 0} \sqrt{\sum_{i=\cstr+1}^n\tilh_i^2-\frac{((\tilh^T\z)-\sum_{i=1}^{\cstr}\tilh_i)^2}{n-\cstr}} p(\h)d\h\leq \int_{\h} \sqrt{\sum_{i=\cstr+1}^n\tilh_i^2-\frac{((\tilh^T\z)-\sum_{i=1}^{\cstr}\tilh_i)^2}{n-\cstr}} p(\h)d\h \nonumber \\
& = & E\sqrt{\sum_{i=\cstr+1}^n\tilh_i^2-\frac{((\tilh^T\z)-\sum_{i=1}^{\cstr}\tilh_i)^2}{n-\cstr}} \leq \sqrt{E\sum_{i=\cstr+1}^n\tilh_i^2-\frac{(E(\tilh^T\z)-E\sum_{i=1}^{\cstr}\tilh_i)^2}{n-\cstr}}. \label{eq:strexp8}
\end{eqnarray}
Finally, the following lemma easily follows by combining (\ref{eq:strexp5}), (\ref{eq:strexp7}), and (\ref{eq:strexp8}).

\begin{lemma}
Assume the setup of Lemma \ref{lm:Bstr}. Let further $\psi_s=\frac{E((\tilh^T\z)-\sum_{i=1}^{\cstr}\tilh_i)}{n}$.Then
\begin{equation}
E(\Bstr)\leq \sqrt{n}\left
(\exp\left \{-\frac{n\epsilon^2\deltastr}{2(1+\epsilon)} \right \}+\exp \left \{  -\frac{\epsilon^2 \psi_s^2 n}{2} \right \}\right )
+\sqrt{E\sum_{i=\cstr+1}^n\tilh_i^2-\frac{(E(\tilh^T\z)-E\sum_{i=1}^{\cstr}\tilh_i)^2}{n-\cstr}}.\label{eq:strexp9}
\end{equation}\label{lm:EBstr}
\end{lemma}
\begin{proof}
Follows from the previous discussion.
\end{proof}

If $n$ is large the first term in (\ref{eq:strexp9}) goes to zero. Then from (\ref{eq:thmesh}), (\ref{eq:strgoal}), and (\ref{eq:strexp9}) it easily follows that for a fixed $\alpha$ one can determine $\beta_s$ as a maximum $\beta$ such that
\begin{equation}
\alpha> \frac{E\sum_{i=\cstr+1}^n\tilh_i^2}{n}-\frac{(E(\tilh^T\z)-E\sum_{i=1}^{\cstr}\tilh_i)^2}{n(n-\cstr)}.\label{eq:stralpha}
\end{equation}
We recall that $k=\beta n$ and $\z\in R^n$ is a column vector such that $\z_i=1,1\leq i\leq (n-k)$ and $\z_i=-1,n-k+1\leq i\leq n$. Therefore, in the above equation $\beta$ is hidden in $\z$. It is relatively easy to see that problem of finding $\beta_s$ for a given fixed $\alpha$ is equivalent to finding minimum $\alpha$ such that (\ref{eq:stralpha}) holds for a fixed $\beta_s$. Let $\beta_s^{max}$ be $\beta_s$ such that minimum $\alpha$ that satisfies (\ref{eq:stralpha}) is $1$. Our goal is then to determine minimum $\alpha$ that satisfies (\ref{eq:stralpha}) for any $\beta_s\in[0, \beta_s^{max}]$.

Therefore in the rest of this subsection we show how the left hand side of (\ref{eq:stralpha}) can be computed for a randomly chosen fixed $\beta_s$. We do so in two steps:
\begin{enumerate}
\item We first determine $\cstr$
\item We then compute $\frac{E\sum_{i=\cstr+1}^n\tilh_i^2}{n}-\frac{(E(\tilh^T\z)-E\sum_{i=1}^{\cstr}\tilh_i)^2}{n(n-\cstr)}$ with $\cstr$ found
in step $1$.
\end{enumerate}

\underline{\emph{Step 1:}}

From Lemma \ref{lm:Bstr} we have $\cstr=\deltastr n$ is a $c$ such that
\begin{eqnarray}
& & \frac{(1-\epsilon)E((\sum_{i=1}^{n-\beta_s n}\tilh_i-\sum_{i=n-\beta_s n+1}^n\tilh_i)-\sum_{i=1}^c\tilh_i)}{n-c}-F_a^{-1}\left (\frac{(1+\epsilon)c}{n}\right )= 0\nonumber \\
& \Leftrightarrow & \frac{(1-\epsilon)(E\sum_{i=\deltastr n+1}^{n}\tilh_i-2E\sum_{i=n-\beta_s n+1}^n\tilh_i)}{n(1-\deltastr)}-F_a^{-1}\left (\frac{(1+\epsilon)\deltastr n}{n}\right )= 0\label{eq:strdelta1}
\end{eqnarray}
where as in Lemma \ref{lm:Bstr} $\tilh_i=|\h|_{(i)}$ and $|\h|_{(i)}$ is the $i$-th smallest magnitude of vector $\h$ with i.i.d. zero-mean unit variance Gaussian random variables and $\epsilon>0$ is an arbitrarily small constant. Set $\theta_s=1-\deltastr$. Following \cite{Stigler73,BarvSam} we have
\begin{equation}
\lim_{n\rightarrow\infty}\frac{E\sum_{i=(1-\theta_s) n+1}^{n}\tilh_i}{n}=\int_{F_a^{-1}(1-\theta_s)}^{\infty}tdF_a(t).\label{eq:strdelta2}
\end{equation}
We then easily compute $F_a^{-1}(1-\theta_s)$ in the following way
\begin{eqnarray}
& & \frac{2}{\sqrt{\pi}}\int_0^{F_a^{-1}(1-\theta_s)}e^{-\frac{t^2}{2}}dt=1-\theta_s\nonumber \\
& \Longrightarrow & F_a^{-1}(1-\theta_s)=\sqrt{2}\erfinv(1-\theta_s)\label{eq:strdelta3}
\end{eqnarray}
where $\erfinv$ is the inverse of the standard error function of the normal random variable.
We further find
\begin{equation}
\int_{F_a^{-1}(1-\theta_s)}^{\infty}tdF_a(t)=\sqrt{\frac{2}{\pi}}\int_{F_a^{-1}(1-\theta_s)}^{\infty}te^{-\frac{t^2}{2}}dt=\sqrt{\frac{2}{\pi}}
e^{-(\erfinv(1-\theta_s))^2}.\label{eq:strdelta4}
\end{equation}
Combining (\ref{eq:strdelta2}) and (\ref{eq:strdelta4}) we obtain
\begin{equation}
\lim_{n\rightarrow\infty}\frac{E\sum_{i=(1-\theta_s) n+1}^{n}\tilh_i}{n}=\sqrt{\frac{2}{\pi}}
e^{-(\erfinv(1-\theta_s))^2}.\label{eq:strdelta5}
\end{equation}
In a completely analogous way we obtain
\begin{equation}
\lim_{n\rightarrow\infty}\frac{E\sum_{i=(1-\beta_s) n+1}^{n}\tilh_i}{n}=\sqrt{\frac{2}{\pi}}
e^{-(\erfinv(1-\beta_s))^2}.\label{eq:strdelta6}
\end{equation}
Similarly to (\ref{eq:strdelta3}) we easily determine
\begin{eqnarray}
& & \frac{2}{\sqrt{\pi}}\int_0^{F_a^{-1}\left ( \frac{(1+\epsilon)\deltastr n}{n}\right )}e^{-\frac{t^2}{2}}dt=\left ( \frac{(1+\epsilon)\deltastr n}{n}\right )\nonumber \\
& \Longrightarrow & F_a^{-1}\left ( \frac{(1+\epsilon)\deltastr n}{n}\right )=\sqrt{2}\erfinv\left ( \frac{(1+\epsilon)\deltastr n}{n}\right )
=\sqrt{2}\erfinv ((1+\epsilon)(1-\theta_s)).\label{eq:strdelta7}
\end{eqnarray}
Combination of (\ref{eq:strdelta1}), (\ref{eq:strdelta5}), (\ref{eq:strdelta6}), and (\ref{eq:strdelta7}) gives us the following equation for computing $\theta_s$
\begin{equation}
(1-\epsilon)\frac{\sqrt{\frac{2}{\pi}}e^{-(\erfinv(1-\theta_s))^2}-2\sqrt{\frac{2}{\pi}}e^{-(\erfinv(1-\beta_s))^2}}{\theta_s}-\sqrt{2}\erfinv ((1+\epsilon)(1-\theta_s))=0.\label{eq:strtheta}
\end{equation}
Let $\htheta_s$ be the solution of (\ref{eq:strtheta}). Then $\deltastr=1-\htheta_s$ and $\cstr=\deltastr n=(1-\htheta_s)n$. This concludes step $1$.

\underline{\emph{Step $2$:}}

In this step we compute
$\frac{E\sum_{i=\cstr+1}^n\tilh_i^2}{n}-\frac{(E(\tilh^T\z)-E\sum_{i=1}^{\cstr}\tilh_i)^2}{n(n-\cstr)}$ with $\cstr=(1-\hat{\theta_s})n$.
Using the results from step $1$ we easily find
\begin{equation}
\lim_{n\rightarrow\infty}\frac{(E(\tilh^T\z)-E\sum_{i=1}^{\cstr}\tilh_i)^2}{n(n-\cstr)}=
\frac{\left (\sqrt{\frac{2}{\pi}}e^{-(\erfinv(1-\hat{\theta_s}))^2}-2\sqrt{\frac{2}{\pi}}e^{-(\erfinv(1-\beta_s))^2}\right )^2}{\hat{\theta_s}}.\label{eq:strstep21}
\end{equation}
Effectively, what is left to compute is $\frac{E\sum_{i=\cstr+1}^n\tilh_i^2}{n}$. Using an approach similar to the one from step $1$ and following \cite{Stigler73,BarvSam} we have
\begin{equation}
\lim_{n\rightarrow\infty}\frac{E\sum_{i=(1-\htheta_s) n+1}^{n}\tilh_i^2}{n}=\int_{F_b^{-1}(1-\htheta_s)}^{\infty}tdF_b(t)\label{eq:strstep22}
\end{equation}
where $F_b^{-1}$ is the inverse cdf of the squared zero-mean unit variance Gaussian random variable.
We then easily compute $F_b^{-1}(1-\htheta_s)$ in the following way
\begin{eqnarray}
& & \frac{1}{\sqrt{2\pi}}\int_0^{F_b^{-1}(1-\htheta_s)}\frac{e^{-\frac{t^2}{2}}}{\sqrt{t}}dt=1-\htheta_s\nonumber \\
& \Longrightarrow & F_b^{-1}(1-\htheta_s)=2(\erfinv(1-\htheta_s))^2.\label{eq:strstep23}
\end{eqnarray}
We further find
\begin{equation}
\int_{F_b^{-1}(1-\htheta_s)}^{\infty}tdF_b(t)=\sqrt{\frac{1}{2\pi}}\int_{F_b^{-1}(1-\htheta_s)}^{\infty}\sqrt{t}e^{-\frac{t^2}{2}}dt
=\frac{1}{\sqrt{2\pi}}\left (\sqrt{2\pi}+2\frac{\sqrt{F_b^{-1}(1-\htheta_s)}}{\exp\left \{ \frac{F_b^{-1}(1-\htheta_s)}{2}\right \}}-\sqrt{2\pi}(1-\htheta_s)\right ).\label{eq:strstep24}
\end{equation}
Combining (\ref{eq:strstep22}) and (\ref{eq:strstep24}) we obtain
\begin{eqnarray}
\lim_{n\rightarrow\infty}\frac{E\sum_{i=(1-\htheta_s) n+1}^{n}\tilh_i^2}{n} & = &
\frac{1}{\sqrt{2\pi}}\left (\sqrt{2\pi}+2\frac{\sqrt{F_b^{-1}(1-\htheta_s)}}{\exp\left \{ \frac{F_b^{-1}(1-\htheta_s)}{2}\right \}}-\sqrt{2\pi}(1-\htheta_s)\right )\nonumber \\
& = & \frac{1}{\sqrt{2\pi}}\left (\sqrt{2\pi}+2\frac{\sqrt{2(\erfinv(1-\htheta_s))^2}}{e^{(\erfinv(1-\htheta_s))^2}}-\sqrt{2\pi}(1-\htheta_s)\right ).\label{eq:strstep25}
\end{eqnarray}

We summarize the results from this section in the following theorem.

\begin{theorem}(Strong threshold)
Let $A$ be an $m\times n$ measurement matrix in (\ref{eq:system})
with the null-space uniformly distributed in the Grassmanian. Let
the unknown $\x$ in (\ref{eq:system}) be $k$-sparse. Let $k,m,n$ be large
and let $\alpha=\frac{m}{n}$ and $\beta_s=\frac{k}{n}$ be constants
independent of $m$ and $n$. Let $\erfinv$ be the inverse of the standard error function associated with zero-mean unit variance Gaussian random variable. Further,
let $\epsilon>0$ be an arbitrarily small constant and $\htheta_s$, ($\beta_s\leq \htheta_s\leq 1$) be the solution of
\begin{equation}
(1-\epsilon)\frac{\sqrt{\frac{2}{\pi}}e^{-(\erfinv(1-\theta_s))^2}-2\sqrt{\frac{2}{\pi}}e^{-(\erfinv(1-\beta_s))^2}}{\theta_s}-\sqrt{2}\erfinv ((1+\epsilon)(1-\theta_s))=0.\label{eq:thmstrtheta}
\end{equation}
If $\alpha$ and $\beta_s$ further satisfy
\begin{equation}
\alpha>\frac{1}{\sqrt{2\pi}}\left (\sqrt{2\pi}+2\frac{\sqrt{2(\erfinv(1-\htheta_s))^2}}{e^{(\erfinv(1-\htheta))^2}}-\sqrt{2\pi}(1-\htheta_s)\right )
-\frac{\left (\sqrt{\frac{2}{\pi}}e^{-(\erfinv(1-\htheta_s))^2}-2\sqrt{\frac{2}{\pi}}e^{-(\erfinv(1-\beta_s))^2}\right )^2}{\htheta_s}\label{eq:thmstralpha}
\end{equation}
then the solutions of (\ref{eq:system}) and (\ref{eq:l1}) coincide with overwhelming
probability.\label{thm:thmstrthr}
\end{theorem}
\begin{proof}
Follows from the previous discussion combining (\ref{eq:thmesh}), (\ref{eq:strgoal}), (\ref{eq:stronglemmaub}), (\ref{eq:strexp9}), (\ref{eq:stralpha}), (\ref{eq:strtheta}), (\ref{eq:strstep21}), and (\ref{eq:strstep25}).
\end{proof}

\begin{figure}[htb]
%%%%%\begin{minipage}[b]{1.0\linewidth}
\centering
\centerline{\epsfig{figure=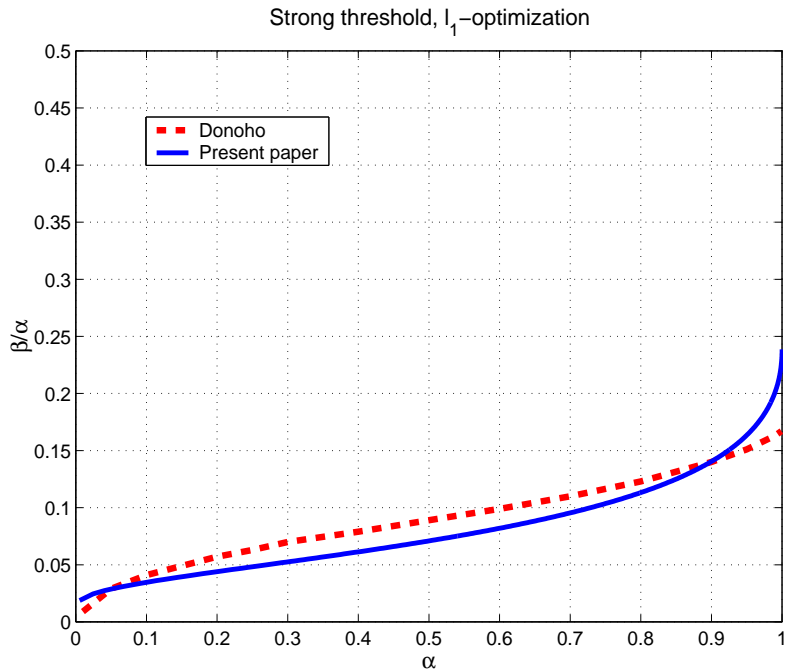,width=10.5cm,height=9cm}}
%%%%%%\end{minipage}
\caption{\emph{Strong} threshold, $\ell_1$-optimization}
\label{fig:strong}
\end{figure}

The results for the strong threshold obtained from the above theorem as well as the best currently known ones from \cite{DonohoPol,DonohoUnsigned}
are presented on Figure \ref{fig:strong}. As can be seen, the threshold results obtained from the previous analysis are comparable to those from \cite{DonohoPol,DonohoUnsigned}
in a large portion of the range for $\alpha$. For the values of $\alpha$ that are close to $1$ the threshold values from Theorem \ref{thm:thmstrthr} are slightly better than those from \cite{DonohoPol,DonohoUnsigned}. For $\alpha\longrightarrow 1$ we have $\beta\approx .24$ which matches the value obtained in \cite{DMK07} and is in fact optimal.

%%%%%%%%%%%%%%%%%%%%%%%%%%%%%%%%%%%%%%%%%%%%%%%%%%%%%%%%%%%
\subsection{Weak threshold} \label{sec:weakgen}
%%%%%%%%%%%%%%%%%%%%%%%%%%%%%%%%%%%%%%%%%%%%%%%%%%%%%%%%%%%

In this subsection we determine the weak threshold $\beta_w$. Before proceeding further we quickly recall on the definition of the weak threshold. Namely, for a given $\alpha$, $\beta_w$ is the maximum value of $\beta$ such that the solutions of (\ref{eq:system}) and (\ref{eq:l1}) coincide for any given $\beta n$-sparse $\x$ with a fixed location of nonzero components and a fixed combination of signs of its elements. Since the analysis will clearly be irrelevant with respect to what particular location and what particular combination of signs of nonzero elements are chosen, we can for the simplicity of the exposition and without loss of generality assume that the components $\x_{1},\x_{2},\dots,\x_{n-k}$ of $\x$ are equal to zero and the components $\x_{n-k+1},\x_{n-k+2},\dots,\x_n$ of $\x$ are smaller than or equal to zero. Under this assumption we have the following corollary of Theorem \ref{thm:thmgen}.
\begin{corollary}(Nonzero part of $\x$ has fixed signs and location)
Assume that an $m\times n$ measurement matrix $A$ is given. Let $\x$
be a $k$-sparse vector whose nonzero components are negative. Also let $\x_1=\x_2=\dots=\x_{n-k}=0.$
Further, assume that $\y=A\x$ and that $\w$ is
an $n\times 1$ vector. Then (\ref{eq:l1}) will
produce the solution of (\ref{eq:system}) if
\begin{equation}
(\forall \w\in \textbf{R}^n | A\w=0) \quad  \sum_{i=n-k+1}^n \w_i<\sum_{i=1}^{n-k}|\w_{i}|.
\label{eq:thmeqgenweak}
\end{equation}
\label{thm:thmgenweak}
\end{corollary}
Following the procedure of Subsection \ref{sec:stronggen} we set
$\Sw$
\begin{equation}
\Sw=\{\w\in S^{n-1}| \quad \sum_{i=n-k+1}^n \w_i<\sum_{i=1}^{n-k}|\w_{i}|\}\label{eq:defSw}
\end{equation}
and
\begin{equation}
w(\Sw)=E\sup_{\w\in \Sw} (\h^T\w) \label{eq:widthdefSw}
\end{equation}
where as earlier $\h$ is a random column vector in $R^{n}$ with i.i.d. ${\cal
N}(0,1)$ components and $S^{n-1}$ is the unit $n$-dimensional sphere. As in Subsection \ref{sec:stronggen} our goal will be to compute an upper bound on $w(\Sw)$ and then equal that upper bound to $\left (\sqrt{m}-\frac{1}{4\sqrt{m}}\right )$. In the following subsections we present a way to get such an upper bound. As earlier, to simplify the exposition we again set $w(\h,\Sw)=\max_{\w\in \Sw} (\h^T\w)$. In order to upper-bound $w(\Sw)$ we will first in Subsection \ref{sec:weakub} determine an upper bound $\Bweak$ on $w(\h,\Sw)$. The expected value with respect to $\h$ of such an upper bound will be an upper bound on $w(\Sw)$. In Subsection \ref{sec:weakexp} we will compute an upper bound on that expected value, i.e. we will compute an upper bound on $E (\Bweak)$. That quantity will be an upper bound on $w(\Sw)$ since according to the following $E(\Bweak)$ is an upper bound on $w(\Sw)$
\begin{equation}
w(\Sw)=Ew(\h,\Sw)=E(\max_{\w\in \Sw} (\h^T\w))\leq E(\Bweak).\label{eq:weakgoal}
\end{equation}

%%%%%%%%%%%%%%%%%%%%%%%%%%%%%%%%%%%%%%%%%%%%%%%%%%%%%%%%%%%%%%
\subsubsection{Upper-bounding $w(\h,\Sw)$}
\label{sec:weakub}
%%%%%%%%%%%%%%%%%%%%%%%%%%%%%%%%%%%%%%%%%%%%%%%%%%%%%%%%%%%%%%%

As in (\ref{eq:workws0}) we have
\begin{equation}
w(\h,\Sw)=\max_{\w\in \Sw} (\h^T\w)=\max_{\w\in \Sw} (\sum_{i=1}^{n-k} |\h_i\w_i|+\sum_{i=n-k+1}^n\h_i\w_i)=\max_{\w\in \Sw} (\sum_{i=1}^{n-k} |\h_i||\w_i|+\sum_{i=n-k+1}^n\h_i\w_i).\label{eq:workww0}
\end{equation}
Let $\h_{1:(n-k)}=(\h_1,\h_2,\dots,\h_{n-k})^T$. Further, let now
$|\h|_{(i)}^{(n-k)}$ be the $i$-th smallest magnitude of elements of $\h_{1:(n-k)}$. Set
\begin{equation}
\hw=(|\h|_{(1)}^{(n-k)},|\h|_{(2)}^{(n-k)},\dots,|\h|_{(n-k)}^{(n-k)},\h_{n-k+1},\h_{n-k+2},\dots,\h_{n})^T.\label{eq:defhw}
\end{equation}
Then one can simplify (\ref{eq:workww0}) in the following way
\begin{eqnarray}
w(\h,\Sw) = \max_{\y\in R^n} & &  \hw^T\y_i \nonumber \\
\mbox{subject to} &  & \y_i\geq 0, 0\leq i\leq (n-k)\nonumber \\
& & \sum_{i=n-k+1}^n\y_i\geq \sum_{i=1}^{n-k} \y_i \nonumber \\
& & \sum_{i=1}^n\y_i^2\leq 1.\label{eq:workww1}
\end{eqnarray}
One can then proceed in a fashion similar to the one from Subsection \ref{sec:strongub} and compute an upper bound based on duality. The only differences are that we now have $\hw$ instead of $\tilh$ and the positive components of $\y$ are only those with indexes less than or equal to $(n-k)$. After repeating literally every step of the derivation from Subsection \ref{sec:strongub} one obtains the following analogue to the equation (\ref{eq:strfinalub})
\begin{equation}
w(\h,\Sw)\leq\sqrt{\sum_{i=1}^n\hw_i^2-\sum_{i=1}^{c}\hw_i^2-\frac{((\hw^T\z)-\sum_{i=1}^{c}\hw_i)^2}{n-c}}
=\sqrt{\sum_{i=c+1}^n\hw_i^2-\frac{((\hw
^T\z)-\sum_{i=1}^{c}\hw_i)^2}{n-c
}}\label{eq:weakfinalub}
\end{equation}
where $\hw_i$ is the $i$-th element of $\hw$ and $c\leq (n-k)$ is such that $((\hw
^T\z)-\sum_{i=1}^{c}\hw_i)\geq 0$. Clearly, as long as $(\hw^T\z)\geq 0$ there will be a $c$ (it is possible that $c=0$) such that quantity on the most right hand side of
(\ref{eq:weakfinalub}) is an upper bound on $w(\h,\Sw)$.

Using (\ref{eq:weakfinalub}) we then establish the following analogue to Lemma \ref{lm:Bstr}.
\begin{lemma}
Let $\h\in R^n$ be a vector with i.i.d. zero-mean unit variance gaussian components. Further let $\hw$ be as defined in (\ref{eq:defhw}) and $w(\h,\Sw)=\max_{\w\in \Sw} (\h^T\w)$ where $\Sw$ is as defined in (\ref{eq:defSw}). Let $\z\in R^n$ be a column vector such that $\z_i=1,1\leq i\leq (n-k)$ and $\z_i=-1,n-k+1\leq i\leq n$. Then
\begin{equation}
w(\h,\Sw)\leq \Bweak \label{eq:weaklemmaub}
\end{equation}
where
\begin{equation}
\Bweak=
\begin{cases}
\sqrt{\sum_{i=1}^n\hw_i^2} & \mbox{if} \quad \zeta_w(\h,\cweak)\leq 0 \\
\sqrt{\sum_{i=\cweak+1}^n\hw_i^2-\frac{((\hw^T\z)-\sum_{i=1}^{\cweak}\hw_i)^2}{n-\cweak}} & \mbox{if} \quad \zeta_w(\h,\cweak)>0
\end{cases},
\end{equation}
$\zeta_w(\h,c)=\frac{(\hw^T\z)-\sum_{i=1}^c\hw_{i}}{n-c}-\hw_{c}$
and $\cweak=\delta_w n$ is a $c\leq n-k$ such that
\begin{equation}
\frac{(1-\epsilon)E((\hw^T\z)-\sum_{i=1}^c\hw_i)}{n-c}-F_a^{-1}\left (\frac{(1+\epsilon)c}{n(1-\beta_w)}\right )= 0.
\end{equation}
$F_a^{-1}(\cdot)$ is the inverse cdf of the random variable $|X|$ where $X$ is zero-mean, unit variance gaussian random variable. $\epsilon>0$ is an arbitrarily small constant independent of $n$.\label{lm:Bweak}
\end{lemma}
\begin{proof}
Follows directly from the derivation before Lemma \ref{lm:Bstr} by replacing $\tilh$ by $\hw$ and by noting that we now have $n(1-\beta_w)$ sorted magnitudes instead of $n$.
\end{proof}

%%%%%%%%%%%%%%%%%%%%%%%%%%%%%%%%%%%%%%%%%%%%%
\subsubsection{Computing  an upper bound on $E(\Bweak)$} \label{sec:weakexp}
%%%%%%%%%%%%%%%%%%%%%%%%%%%%%%%%%%%%%%%%%%%%%

Following step-by-step the derivation of Lemma \ref{lm:EBstr} (with a trivial adjustment in computing Lipschitz constant $\sigma$) we can establish the weak threshold analogue to it.
\begin{lemma}
Assume the setup of Lemma \ref{lm:Bweak}. Let further $\psi_w=\frac{E(\hw^T\z)-\sum_{i=1}^{\cweak}\hw_i)}{n}$.Then
\begin{equation}
E(\Bweak)\leq \sqrt{n}\left
(\exp\left \{-\frac{n\epsilon^2\delta_w}{2(1+\epsilon)} \right \}+\exp \left \{  -\frac{\epsilon^2 \psi_w^2 n}{2} \right \}\right )
+\sqrt{E\sum_{i=\cweak+1}^n\hw_i^2-\frac{(E(\hw^T\z)-E\sum_{i=1}^{\cweak}\hw_i)^2}{n-\cweak}}.\label{eq:weakexp9}
\end{equation}\label{lm:EBweak}
\end{lemma}
\begin{proof}
Follows directly from the derivation before Lemma \ref{lm:EBstr}.
\end{proof}

As in (\ref{eq:stralpha}), if $n$ is large, for a fixed $\alpha$ one can determine $\beta_w$ as a maximum $\beta$ such that
\begin{equation}
\alpha\geq \frac{E\sum_{i=\cweak+1}^n\hw_i^2}{n}-\frac{(E(\hw^T\z)-E\sum_{i=1}^{\cweak}\hw_i)^2}{n(n-\cweak)}.\label{eq:weakalpha}
\end{equation}
 As earlier $k=\beta n$ and $\z\in R^n$ is a column vector such that $\z_i=1,1\leq i\leq (n-k)$ and $\z_i=-1,n-k+1\leq i\leq n$. Also, as in Subsection \ref{sec:strongexp}, $\beta$ is again hidden in $\z$. It is not difficult to see that problem of finding $\beta_w$ for a given fixed $\alpha$ is equivalent to finding minimum $\alpha$ such that (\ref{eq:weakalpha}) holds for a fixed $\beta_w$. Let $\beta_w^{max}$ be $\beta_w$ such that minimum $\alpha$ that satisfies (\ref{eq:weakalpha}) is $1$. Analogously to what was done in Subsection \ref{sec:strongexp}, we will determine minimum $\alpha$ that satisfies (\ref{eq:weakalpha}) for any $\beta_w\in[0, \beta_w^{max}]$.

Therefore in the rest of this subsection we show how the left hand side of (\ref{eq:weakalpha}) can be computed for a randomly chosen fixed $\beta_w$. We, as in as in Subsection \ref{sec:strongexp}, do so in two steps:
\begin{enumerate}
\item We first determine $\cweak$
\item We then compute $\frac{E\sum_{i=\cweak+1}^n\hw_i^2}{n}-\frac{(E(\hw^T\z)-E\sum_{i=1}^{\cweak}\hw_i)^2}{n(n-\cweak)}$ with $\cweak$ found
in step $1$.
\end{enumerate}

\underline{\emph{Step 1:}}

From Lemma \ref{lm:Bweak} we have $\cweak=\delta_w n$ is a $c$ such that
\begin{eqnarray}
& & \frac{(1-\epsilon)E((\sum_{i=1}^{n-\beta_w n}\hw_{i}-\sum_{i=n-\beta_w n+1}^n\hw_{i})-\sum_{i=1}^{\delta_w n}\hw_i)}{n-c}-F_a^{-1}\left (\frac{(1+\epsilon)c}{n(1-\beta_w)}\right )= 0\nonumber \\
& \Leftrightarrow & \frac{(1-\epsilon)(E\sum_{i=1}^{n-\beta_w n}\hw_i-E\sum_{i=n-\beta_w n+1}^n\h_{i}-E\sum_{i=1}^{\delta_w n}\hw_i)}{n-c}-F_a^{-1}\left (\frac{(1+\epsilon)c}{n(1-\beta_w)}\right )= 0\label{eq:weakdelta1}
\end{eqnarray}
where we recall $\hw_i,1\leq i\leq (n-\beta_w n)$, is the $i$-th smallest magnitude of vector $\h_{1:(n-\beta_w n)}$. We also recall that $\h_{1:(n-\beta_w n)}$ stands for the first $(n-\beta_w n)$ components of $\h$ and $\h_i,n-\beta_w n+1\leq i\leq n$, are naturally the last $\beta_w n$ components of vector $\h$. Also, as always, all components of $\h$ are i.i.d. zero-mean unit variance Gaussian random variables and $\epsilon>0$ is an arbitrarily small constant. Then clearly
$E\h_i=0,n-\beta_w n+1\leq i\leq n$ and we have from (\ref{eq:weakdelta1})
\begin{eqnarray}
& & \frac{(1-\epsilon)E((\sum_{i=1}^{n-\beta_w n}\hw_{i}-\sum_{i=n-\beta_w n+1}^n\hw_{i})-\sum_{i=1}^{\delta_w n}\hw_i)}{n-c}-F_a^{-1}\left (\frac{(1+\epsilon)c}{n(1-\beta_w)}\right )= 0\nonumber \\
& \Leftrightarrow & \frac{(1-\epsilon)E\sum_{i=\delta_w n+1}^{n-\beta_w n}\hw_i}{n(1-\delta_w)}-F_a^{-1}\left (\frac{(1+\epsilon)\delta_w n}{n(1-\beta_w)}\right )= 0.\label{eq:weakdelta2}
\end{eqnarray}
Set $\theta_w=1-\delta_w$. Following \cite{Stigler73,BarvSam} and in a way completely analogous to (\ref{eq:strdelta5}) we obtain
\begin{equation}
\lim_{n\rightarrow\infty}\frac{E\sum_{i=(1-\theta_w) n+1}^{(1-\beta_w)n}\hw_i}{n(1-\beta_w)}=\sqrt{\frac{2}{\pi}}
e^{-(\erfinv(\frac{1-\theta_w}{1-\beta_w}))^2}.\label{eq:weakdelta5}
\end{equation}
As in (\ref{eq:strdelta7}) we have
\begin{eqnarray}
& & \frac{2}{\sqrt{\pi}}\int_0^{F_a^{-1}\left ( \frac{(1+\epsilon)\delta_w n}{n(1-\beta_w)}\right )}e^{-\frac{t^2}{2}}dt=\left ( \frac{(1+\epsilon)\delta_w n}{n(1-\beta_w)}\right )\nonumber \\
& \Longrightarrow & F_a^{-1}\left ( \frac{(1+\epsilon)\delta_w n}{n(1-\beta_w)}\right )=\sqrt{2}\erfinv\left ( \frac{(1+\epsilon)\delta n}{n(1-\beta_w)}\right )
=\sqrt{2}\erfinv ((1+\epsilon)(\frac{1-\theta_w}{1-\beta_w})).\label{eq:weakdelta7}
\end{eqnarray}
Combination of (\ref{eq:weakdelta2}), (\ref{eq:weakdelta5}), and (\ref{eq:weakdelta7}) gives us the following equation for computing $\theta_w$
\begin{equation}
(1-\epsilon)(1-\beta_w)\frac{\sqrt{\frac{2}{\pi}}e^{-(\erfinv(\frac{1-\theta_w}{1-\beta_w}))^2}}{\theta_w}-\sqrt{2}\erfinv ((1+\epsilon)\frac{1-\theta_w}{1-\beta_w})=0.\label{eq:weaktheta}
\end{equation}
Let $\hat{\theta}_w$ be the solution of (\ref{eq:weaktheta}). Then $\delta_w=1-\hat{\theta}_w$ and $\cweak=\delta_w n=(1-\hat{\theta}_w)n$. This concludes step $1$.

\underline{\emph{Step $2$:}}

In this step we compute
$\frac{E\sum_{i=\cweak+1}^n\hw_i^2}{n}-\frac{(E(\hw^T\z)-E\sum_{i=1}^{\cweak}\hw_i)^2}{n(n-\cweak)}$ with $\cweak=(1-\hat{\theta}_w)n$.
Using the results from step $1$ we easily find
\begin{equation}
\lim_{n\rightarrow\infty}\frac{(E(\hw^T\z)-E\sum_{i=1}^{\cweak}\hw_i)^2}{n(n-\cweak)}=
\frac{\left ((1-\beta_w)\sqrt{\frac{2}{\pi}}e^{-(\erfinv(\frac{1-\hat{\theta}_w}{1-\beta_w}))^2}\right )^2}{\hat{\theta}_w}.\label{eq:weakstep21}
\end{equation}
Effectively, what is left to compute is $\frac{E\sum_{i=\cweak+1}^n\hw_i^2}{n}$. First we note that
\begin{equation}
\frac{E\sum_{i=\cweak+1}^n\hw_i^2}{n}=\frac{E\sum_{i=(1-\htheta_w)n+1}^{(1-\beta_w)n}\hw_i^2+E\sum_{i=(1-\beta_w)n+1}^{n}\h_{i}^2}{n}
=\frac{E\sum_{i=(1-\htheta_w)n+1}^{(1-\beta_w)n}\hw_i^2}{n}
+\beta_w.\label{eq:weakstep22}
\end{equation}
Using an approach similar to the one from step $2$ of Subsection \ref{sec:strongexp} and following \cite{Stigler73,BarvSam} we have
\begin{equation}
\lim_{n\rightarrow\infty}\frac{E\sum_{i=(1-\htheta_w)n+1}^{(1-\beta_w)n}\tilh_i^2}{n(1-\beta_w)}=
\int_{F_b^{-1}(\frac{1-\htheta_w}{1-\beta_w})}^{\infty}tdF_b(t)\label{eq:weakstep23}
\end{equation}
where as in Subsection \ref{sec:strongexp} $F_b^{-1}$ is the inverse cdf of squared zero-mean unit variance Gaussian random variable.
Following (\ref{eq:strstep23}) we then have
\begin{eqnarray}
F_b^{-1}(\frac{1-\htheta_w}{1-\beta_w})=2(\erfinv(\frac{1-\htheta_w}{1-\beta_w}))^2.\label{eq:weakstep24}
\end{eqnarray}
As in (\ref{eq:strstep24}) we also find
\begin{equation}
\int_{F_b^{-1}(\frac{1-\htheta_w}{1-\beta})}^{\infty}tdF_b(t)=\frac{1}{\sqrt{2\pi}}\left (\sqrt{2\pi}+2\frac{\sqrt{F_b^{-1}(\frac{1-\htheta_w}{1-\beta_w})}}{\exp\left \{ \frac{F_b^{-1}(\frac{1-\htheta_w}{1-\beta_w})}{2}\right \}}-\sqrt{2\pi}\frac{1-\htheta_w}{1-\beta_w}\right ).\label{eq:weakstep25}
\end{equation}
Combining (\ref{eq:weakstep22}), (\ref{eq:weakstep23}), (\ref{eq:weakstep24}), and (\ref{eq:weakstep25}) we obtain
\begin{eqnarray}
\lim_{n\rightarrow\infty}\frac{E\sum_{i=(1-\htheta_w) n+1}^{n}\hw_i^2}{n}  =  \frac{1-\beta_w}{\sqrt{2\pi}}\left (\sqrt{2\pi}+2\frac{\sqrt{2(\erfinv(\frac{1-\htheta_w}{1-\beta_w}))^2}}{e^{(\erfinv(\frac{1-\htheta_w}{1-\beta_w}))^2}}-\sqrt{2\pi}
\frac{1-\htheta_w}{1-\beta_w}\right )+\beta_w.\label{eq:weakstep26}
\end{eqnarray}

We summarize the results from this section in the following theorem.

\begin{theorem}(Weak threshold)
Let $A$ be an $m\times n$ measurement matrix in (\ref{eq:system})
with the null-space uniformly distributed in the Grassmanian. Let
the unknown $\x$ in (\ref{eq:system}) be $k$-sparse. Further, let the location and signs of nonzero elements of $\x$ be arbitrarily chosen but fixed.
Let $k,m,n$ be large
and let $\alpha=\frac{m}{n}$ and $\beta_w=\frac{k}{n}$ be constants
independent of $m$ and $n$. Let $\erfinv$ be the inverse of the standard error function associated with zero-mean unit variance Gaussian random variable.  Further,
let $\epsilon>0$ be an arbitrarily small constant and $\htheta_w$, ($\beta_w\leq \htheta_w\leq 1$) be the solution of
\begin{equation}
(1-\epsilon)(1-\beta_w)\frac{\sqrt{\frac{2}{\pi}}e^{-(\erfinv(\frac{1-\theta_w}{1-\beta_w}))^2}}{\theta_w}-\sqrt{2}\erfinv ((1+\epsilon)\frac{1-\theta_w}{1-\beta_w})=0.\label{eq:thmweaktheta}
\end{equation}
If $\alpha$ and $\beta_w$ further satisfy
\begin{equation}
\alpha>\frac{1-\beta_w}{\sqrt{2\pi}}\left (\sqrt{2\pi}+2\frac{\sqrt{2(\erfinv(\frac{1-\htheta_w}{1-\beta_w}))^2}}{e^{(\erfinv(\frac{1-\htheta_w}{1-\beta_w}))^2}}-\sqrt{2\pi}
\frac{1-\htheta_w}{1-\beta_w}\right )+\beta_w
-\frac{\left ((1-\beta_w)\sqrt{\frac{2}{\pi}}e^{-(\erfinv(\frac{1-\hat{\theta}_w}{1-\beta_w}))^2}\right )^2}{\hat{\theta}_w}\label{eq:thmweakalpha}
\end{equation}
then the solutions of (\ref{eq:system}) and (\ref{eq:l1}) coincide with overwhelming
probability.\label{thm:thmweakthr}
\end{theorem}
\begin{proof}
Follows from the previous discussion combining (\ref{eq:thmesh}), (\ref{eq:weakgoal}), (\ref{eq:weaklemmaub}), (\ref{eq:weakexp9}), (\ref{eq:weakalpha}), (\ref{eq:weaktheta}), (\ref{eq:weakstep21}), and (\ref{eq:weakstep26}).
\end{proof}

\begin{figure}[htb]
%%%%%\begin{minipage}[b]{1.0\linewidth}
\centering
\centerline{\epsfig{figure=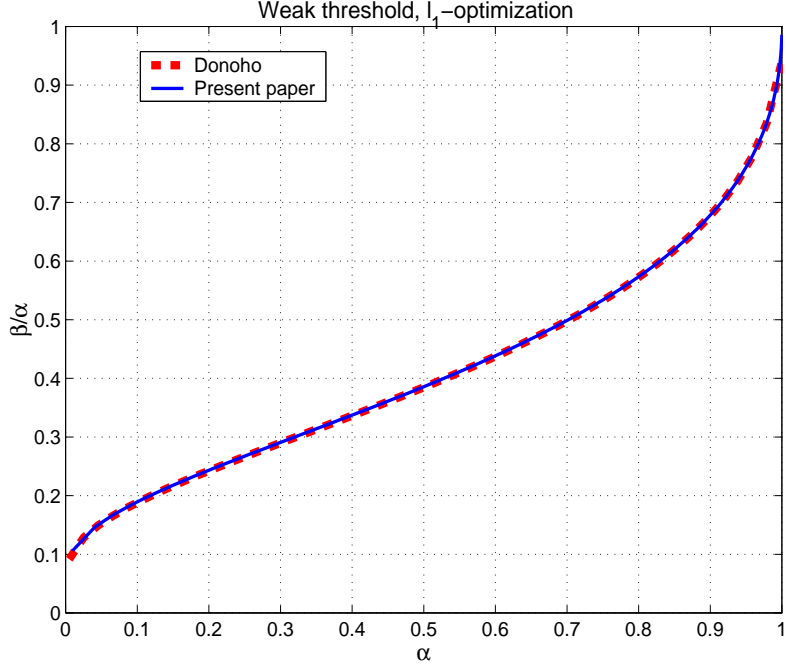,width=10.5cm,height=9cm}}
%%%%%%\end{minipage}
\caption{\emph{Weak} threshold, $\ell_1$-optimization}
\label{fig:weak}
\end{figure}

The results for the weak threshold obtained from the above theorem as well as the best currently known ones from \cite{DonohoPol,DonohoUnsigned}
are presented on Figure \ref{fig:weak}. As can be seen, the threshold results obtained from the previous analysis match those from \cite{DonohoPol,DonohoUnsigned}.

%%%%%%%%%%%%%%%%%%%%%%%%%%%%%%%%%%%%%%%%%%%%%%%%%%%%%%%%%%%
\subsection{Sectional threshold} \label{sec:secgen}
%%%%%%%%%%%%%%%%%%%%%%%%%%%%%%%%%%%%%%%%%%%%%%%%%%%%%%%%%%%

In this subsection we determine the sectional threshold $\betasec$. Before proceeding further we one more time quickly recall on the definition of the sectional threshold. Namely, for a given $\alpha$, $\betasec$ is the maximum value of $\beta$ such that the solutions of (\ref{eq:system}) and (\ref{eq:l1}) coincide for any given $\beta n$-sparse $\x$ with a fixed location of nonzero components. Since the analysis will clearly be irrelevant with respect to what particular location of nonzero elements is chosen, we can for the simplicity of the exposition and without loss of generality assume that the components $\x_{1},\x_{2},\dots,\x_{n-k}$ of $\x$ are equal to zero. Under this assumption we have the following corollary of Theorem \ref{thm:thmgen}.
\begin{corollary}[Nonzero part of $\x$ has a fixed location]
Assume that an $m\times n$ measurement matrix $A$ is given. Let $\x$
be a $k$-sparse vector. Also let $\x_1=\x_2=\dots=\x_{n-k}=0.$
Further, assume that $\y=A\x$ and that $\w$ is
an $n\times 1$ vector. Then (\ref{eq:l1}) will
produce the solution of (\ref{eq:system}) if
\begin{equation}
(\forall \w\in \textbf{R}^n | A\w=0) \quad  \sum_{i=n-k+1}^n |\w_i|<\sum_{i=1}^{n-k}|\w_{i}|.
\label{eq:thmeqgensec}
\end{equation}\label{thm:thmgensec}
\end{corollary}
Following the procedure of Subsection \ref{sec:weakgen} we set
$\Ssec$
\begin{equation}
\Ssec=\{\w\in S^{n-1}| \quad \sum_{i=n-k+1}^n |\w_i|<\sum_{i=1}^{n-k}|\w_{i}|\}\label{eq:defSsec}
\end{equation}
and
\begin{equation}
w(\Ssec)=E\sup_{\w\in \Ssec} (\h^T\w) \label{eq:widthdefSsec}
\end{equation}
where as earlier $\h$ is a random column vector in $R^{n}$ with i.i.d. ${\cal
N}(0,1)$ components and $S^{n-1}$ is the unit $n$-dimensional sphere. As in Subsections \ref{sec:stronggen} and \ref{sec:weakgen} our goal will be to compute an upper bound on $w(\Ssec)$ and then equal that upper bound to $\left ( \sqrt{m}-\frac{1}{4\sqrt{m}}\right )$. In the following subsections we present a way to get such an upper bound. As earlier, we set $w(\h,\Ssec)=\max_{\w\in \Ssec} (\h^T\w)$. Following the strategy of previous sections in Subsection \ref{sec:secub} we determine an upper bound $\Bsec$ on $w(\h,\Ssec)$. In Subsection \ref{sec:secexp} we will compute an upper bound on $E (\Bsec)$. That quantity will be an upper bound on $w(\Ssec)$ since according to the following $E(\Bsec)$ is an upper bound on $w(\Ssec)$
\begin{equation}
w(\Ssec)=Ew(\h,\Ssec)=E(\max_{\w\in \Ssec} (\h^T\w))\leq E(\Bsec).\label{eq:secgoal}
\end{equation}

%%%%%%%%%%%%%%%%%%%%%%%%%%%%%%%%%%%%%%%%%%%%%%%%%%%%%%%%%%%%%%
\subsubsection{Upper-bounding $w(\h,\Ssec)$}
\label{sec:secub}
%%%%%%%%%%%%%%%%%%%%%%%%%%%%%%%%%%%%%%%%%%%%%%%%%%%%%%%%%%%%%%%

Following (\ref{eq:workww0}) we have
\begin{equation}
w(\h,\Ssec)=\max_{\w\in \Ssec} (\h^T\w)=\max_{\w\in \Ssec} (\sum_{i=1}^{n-k} |\h_i\w_i|+\sum_{i=n-k+1}^n|\h_i\w_i|)=\max_{\w\in \Ssec} (\sum_{i=1}^{n-k} |\h_i||\w_i|+\sum_{i=n-k+1}^n|\h_i||\w_i|).\label{eq:workwsec0}
\end{equation}
As earlier, let $\h_{1:(n-k)}=(\h_1,\h_2,\dots,\h_{n-k})^T$ and let $|\h|_{(i)}^{(n-k)}$ be the $i$-th smallest magnitude of elements of $\h_{1:(n-k)}$. Set
\begin{equation}
\hsec=(|\h|_{(1)}^{(n-k)},|\h|_{(2)}^{(n-k)},\dots,|\h|_{(n-k)}^{(n-k)},|\h_{n-k+1}|,|\h_{n-k+2}|,\dots,|\h_{n}|)^T\label{eq:defhsec}
\end{equation}
where $|\h_i|,n-k+1\leq i \leq n$, is the absolute value (magnitude) of $\h_i,n-k+1\leq i\leq n$.
Then one can simplify (\ref{eq:workwsec0}) in the following way
\begin{eqnarray}
w(\h,\Ssec) = \max_{\y\in R^n} & &  \hsec^T\y_i \nonumber \\
\mbox{subject to} &  & \y_i\geq 0, 0\leq i\leq n\nonumber \\
& & \sum_{i=n-k+1}^n\y_i\geq \sum_{i=1}^{n-k} \y_i \nonumber \\
& & \sum_{i=1}^n\y_i^2\leq 1.\label{eq:workwsec1}
\end{eqnarray}
One can then proceed in the similar fashion as in Subsection \ref{sec:strongub} and compute an upper bound based on duality. The only differences is that we now have $\hsec$ instead of $\tilh$. After repeating literally every step of the derivation from Subsection \ref{sec:strongub} one obtains the following analogue to the equation (\ref{eq:strfinalub})
\begin{equation}
w(\h,\Ssec)\leq\sqrt{\sum_{i=1}^n\hsec_i^2-\sum_{i=1}^{c}\hsec_i^2-\frac{((\hsec^T\z)-\sum_{i=1}^{c}\hsec_i)^2}{n-c}}
=\sqrt{\sum_{i=c+1}^n\hsec_i^2-\frac{((\hsec
^T\z)-\sum_{i=1}^{c}\hsec_i)^2}{n-c
}}\label{eq:secfinalub}
\end{equation}
where $\hsec_i$ is the $i$-th element of $\hsec$ and $c\leq (n-k)$ is such that $((\hsec
^T\z)-\sum_{i=1}^{c}\hsec_i)\geq 0$. As earlier, as long as $(\hsec^T\z)\geq 0$ there will be a $c$ (it is possible that $c=0$) such that quantity on the most right hand side of
(\ref{eq:secfinalub}) is an upper bound on $w(\h,\Ssec)$.

Using (\ref{eq:secfinalub}) we then establish the following analogue to Lemmas \ref{lm:Bstr} and \ref{lm:Bweak}.
\begin{lemma}
Let $\h\in R^n$ be a vector with i.i.d. zero-mean unit variance gaussian components. Further let $\hsec$ be as defined in (\ref{eq:defhsec}) and $w(\h,\Ssec)=\max_{\w\in \Ssec} (\h^T\w)$ where $\Ssec$ is as defined in (\ref{eq:defSsec}). Let $\z\in R^n$ be a column vector such that $\z_i=1,1\leq i\leq (n-k)$ and $\z_i=-1,n-k+1\leq i\leq n$. Then
\begin{equation}
w(\h,\Ssec)\leq \Bsec \label{eq:seclemmaub}
\end{equation}
where
\begin{equation}
\Bsec=
\begin{cases}
\sqrt{\sum_{i=1}^n\hsec_i^2} & \mbox{if} \quad \zeta_{sec}(\h,\csec)\leq 0 \\
\sqrt{\sum_{i=\csec+1}^n\hsec_i^2-\frac{((\hsec^T\z)-\sum_{i=1}^{\csec}\hsec_i)^2}{n-\csec}} & \mbox{if} \quad \zeta_{sec}(\h,\csec)>0
\end{cases},
\end{equation}
$\zeta_{sec}(\h,c)=\frac{(\hsec^T\z)-\sum_{i=1}^c\hsec_i}{n-c}-\hsec_{c}$
and $\csec=\deltasec n$ is a $c\leq n-k$ such that
\begin{equation}
\frac{(1-\epsilon)E((\hsec^T\z)-\sum_{i=1}^c\hsec_i)}{n-c}-F_a^{-1}\left (\frac{(1+\epsilon)c}{n(1-\betasec)}\right )= 0.
\end{equation}
$F_a^{-1}(\cdot)$ is the inverse cdf of the random variable $|X|$ where $X$ is zero-mean, unit variance gaussian random variable. $\epsilon>0$ is an arbitrarily small constant independent of $n$.\label{lm:Bsec}
\end{lemma}
\begin{proof}
Follows directly from the derivation before Lemma \ref{lm:Bstr}.
\end{proof}

%%%%%%%%%%%%%%%%%%%%%%%%%%%%%%%%%%%%%%%%%%%%%
\subsubsection{Computing  an upper bound on $E(\Bsec)$} \label{sec:secexp}
%%%%%%%%%%%%%%%%%%%%%%%%%%%%%%%%%%%%%%%%%%%%%

Following step-by-step the derivation of Lemma \ref{lm:EBstr} (with a trivial adjustment in computing Lipschitz constant $\sigma$) we can establish the sectional threshold analogue to it.
\begin{lemma}
Assume the setup of Lemma \ref{lm:Bsec}. Let further $\psi_{sec}=\frac{E(\hsec^T\z)-\sum_{i=1}^{\csec}\hsec_i)}{n}$.Then
\begin{equation}
E(\Bsec)\leq \sqrt{n}\left
(\exp\left \{-\frac{n\epsilon^2\deltasec}{2(1+\epsilon)} \right \}+\exp \left \{  -\frac{\epsilon^2 \psi_{sec}^2 n}{2} \right \}\right )
+\sqrt{E\sum_{i=\csec+1}^n\hsec_i^2-\frac{(E(\hsec^T\z)-E\sum_{i=1}^{\csec}\hsec_i)^2}{n-\csec}}.\label{eq:secexp9}
\end{equation}\label{lm:EBsec}
\end{lemma}
\begin{proof}
Follows directly from the derivation before Lemma \ref{lm:EBstr}.
\end{proof}

As in (\ref{eq:stralpha}), if $n$ is large, for a fixed $\alpha$ one can determine $\betasec$ as a maximum $\beta$ such that
\begin{equation}
\alpha\geq \frac{E\sum_{i=\csec+1}^n\hsec_i^2}{n}-\frac{(E(\hsec^T\z)-E\sum_{i=1}^{\csec}\hsec_i)^2}{n(n-\csec)}.\label{eq:secalpha}
\end{equation}
In the rest of this subsection we show how the left hand side of (\ref{eq:secalpha}) can be computed for a randomly chosen fixed $\betasec$. We again, as earlier, do so in two steps:
\begin{enumerate}
\item We first determine $\csec$
\item We then compute $\frac{E\sum_{i=\csec+1}^n\hsec_i^2}{n}-\frac{(E(\hsec^T\z)-E\sum_{i=1}^{\csec}\hsec_i)^2}{n(n-\csec)}$ with $\csec$ found
in step $1$.
\end{enumerate}

\underline{\emph{Step 1:}}

From Lemma \ref{lm:Bsec} we have $\csec=\deltasec n$ is a $c$ such that
\begin{eqnarray}
& & \frac{(1-\epsilon)E((\sum_{i=1}^{n-\betasec n}\hsec_{i}-\sum_{i=n-\betasec n+1}^n\hsec_{i})-\sum_{i=1}^{\deltasec n}\hsec_i)}{n-c}-F_a^{-1}\left (\frac{(1+\epsilon)c}{n(1-\betasec)}\right )= 0\nonumber \\
& \Leftrightarrow & \frac{(1-\epsilon)(E\sum_{i=1}^{n-\betasec n}\hsec_i-E\sum_{i=n-\betasec n+1}^n|\h_{i}|-E\sum_{i=1}^{\deltasec n}\hsec_i)}{n-c}-F_a^{-1}\left (\frac{(1+\epsilon)c}{n(1-\betasec)}\right )= 0\nonumber \\\label{eq:secdelta1}
\end{eqnarray}
where as in Subsection \ref{sec:weakgen} $|\h|_{(i)}^{(n-k)},1\leq i\leq (n-\betasec n)$, is the $i$-th smallest magnitude of vector $\h_{1:(n-\betasec n)}$. Furthermore, $\hsec_i=|\h|_{(i)}^{(n-k)}$ and clearly  $\h_{1:(n-\betasec n)}$ stands for first $(n-\betasec n)$ components of $\h$. We also recall that $|\h_i|,n-\betasec n+1\leq i\leq n$, are the magnitudes of the last $\betasec n$ components of vector $\h$ (these magnitudes of last $\betasec n$ components of vector $\h$ are not sorted). As earlier, all components of $\h$ are i.i.d. zero-mean unit variance Gaussian random variables and $\epsilon>0$ is an arbitrarily small constant. Then clearly
$E|\h_i|=\sqrt{\frac{2}{\pi}},n-\betasec n+1\leq i\leq n$, and we have from (\ref{eq:secdelta1})
\begin{eqnarray}
& & \frac{(1-\epsilon)E((\sum_{i=1}^{n-\betasec n}\hsec_{i}-\sum_{i=n-\betasec n+1}^n\hsec_{i})-\sum_{i=1}^{\deltasec n}\hsec_i)}{n-c}-F_a^{-1}\left (\frac{(1+\epsilon)c}{n(1-\betasec)}\right )= 0\nonumber \\
& \Leftrightarrow & \frac{(1-\epsilon)E\sum_{i=\deltasec n+1}^{n-\betasec n}\hsec_i-\sqrt{\frac{2}{\pi}}\betasec n}{n(1-\deltasec)}-F_a^{-1}\left (\frac{(1+\epsilon)\deltasec n}{n(1-\betasec)}\right )= 0.\label{eq:secdelta2}
\end{eqnarray}
Set $\thetasec=1-\deltasec$. Following the derivation of (\ref{eq:weakdelta5}) and (\ref{eq:weakdelta7})
we have the following equation for computing $\thetasec$
\begin{equation}
(1-\epsilon)(1-\betasec)\frac{\sqrt{\frac{2}{\pi}}e^{-(\erfinv(\frac{1-\theta_w}{1-\beta_w}))^2}-\sqrt{\frac{2}{\pi}}\frac{\betasec}{1-\betasec}}
{\thetasec}-\sqrt{2}\erfinv ((1+\epsilon)\frac{1-\thetasec}{1-\betasec})=0.\label{eq:sectheta}
\end{equation}
Let $\hthetasec$ be the solution of (\ref{eq:sectheta}). Then $\deltasec=1-\hthetasec$ and $\csec=\deltasec n=(1-\hthetasec)n$. This concludes step $1$.

\underline{\emph{Step $2$:}}

In this step we compute
$\frac{E\sum_{i=\csec+1}^n\hsec_i^2}{n}-\frac{(E(\hsec^T\z)-E\sum_{i=1}^{\csec}\hsec_i)^2}{n(n-\csec)}$ with $\csec=(1-\hthetasec)n$.
Using results from step $1$ we easily find
\begin{equation}
\lim_{n\rightarrow\infty}\frac{(E(\hsec^T\z)-E\sum_{i=1}^{\csec}\hsec_i)^2}{n(n-\csec)}=
\frac{\left ((1-\betasec)\sqrt{\frac{2}{\pi}}e^{-(\erfinv(\frac{1-\hthetasec}{1-\betasec}))^2}-\betasec\sqrt{\frac{2}{\pi}}\right )^2}{\hthetasec}.\label{eq:secstep21}
\end{equation}
Effectively, what is left to compute is $\frac{E\sum_{i=\csec+1}^n\hsec_i^2}{n}$. However, the same quantity has already been computed in
(\ref{eq:weakstep26}). Hence we have
\begin{equation}
\lim_{n\rightarrow\infty}\frac{E\sum_{i=(1-\hthetasec) n+1}^{n}\hsec_i^2}{n}  =  \frac{1-\betasec}{\sqrt{2\pi}}\left (\sqrt{2\pi}+2\frac{\sqrt{2(\erfinv(\frac{1-\hthetasec}{1-\betasec}))^2}}{e^{(\erfinv(\frac{1-\hthetasec}{1-\betasec}))^2}}-\sqrt{2\pi}
\frac{1-\hthetasec}{1-\betasec}\right )+\betasec.\label{eq:secstep26}
\end{equation}

We summarize the results from this section in the following theorem.

\begin{figure}[htb]
%%%%%\begin{minipage}[b]{1.0\linewidth}
\centering
\centerline{\epsfig{figure=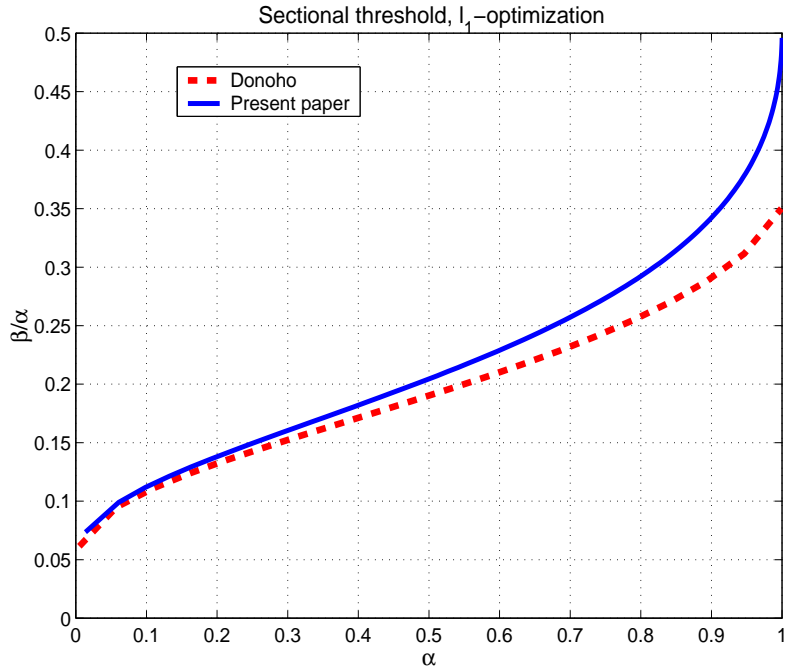,width=10.5cm,height=9cm}}
%%%%%%\end{minipage}
\caption{\emph{Sectional} threshold, $\ell_1$-optimization}
\label{fig:sec}
\end{figure}

\begin{theorem}(Sectional threshold)
Let $A$ be an $m\times n$ measurement matrix in (\ref{eq:system})
with the null-space uniformly distributed in the Grassmanian. Let
the unknown $\x$ in (\ref{eq:system}) be $k$-sparse. Further, let the location of nonzero elements of $\x$ be arbitrarily chosen but fixed.
Let $k,m,n$ be large
and let $\alpha=\frac{m}{n}$ and $\betasec=\frac{k}{n}$ be constants
independent of $m$ and $n$. Let $\erfinv$ be the inverse of the standard error function associated with zero-mean unit variance Gaussian random variable.  Further,
let $\epsilon>0$ be an arbitrarily small constant and $\hthetasec$, ($\betasec\leq \hthetasec\leq 1$) be the solution of
\begin{equation}
(1-\epsilon)(1-\betasec)\frac{\sqrt{\frac{2}{\pi}}e^{-(\erfinv(\frac{1-\thetasec}{1-\betasec}))^2}-\sqrt{\frac{2}{\pi}}\frac{\betasec}{1-\betasec}}
{\thetasec}-\sqrt{2}\erfinv ((1+\epsilon)\frac{1-\thetasec}{1-\betasec})=0.\label{eq:thmsectheta}
\end{equation}
If $\alpha$ and $\betasec$ further satisfy
\begin{equation}
\hspace{-.6in}\alpha>\frac{1-\betasec}{\sqrt{2\pi}}\left (\sqrt{2\pi}+2\frac{\sqrt{2(\erfinv(\frac{1-\hthetasec}{1-\betasec}))^2}}{e^{(\erfinv(\frac{1-\hthetasec}{1-\betasec}))^2}}-\sqrt{2\pi}
\frac{1-\hthetasec}{1-\betasec}\right )+\betasec
-\frac{\left ((1-\betasec)\sqrt{\frac{2}{\pi}}e^{-(\erfinv(\frac{1-\hthetasec}{1-\betasec}))^2}-\sqrt{\frac{2}{\pi}}\betasec\right )^2}{\hthetasec}\label{eq:thmsecalpha}
\end{equation}
then the solutions of (\ref{eq:system}) and (\ref{eq:l1}) coincide with overwhelming
probability.\label{thm:thmsecthr}
\end{theorem}
\begin{proof}
Follows from the previous discussion combining (\ref{eq:thmesh}), (\ref{eq:secgoal}), (\ref{eq:seclemmaub}), (\ref{eq:secexp9}), (\ref{eq:secalpha}), (\ref{eq:sectheta}), (\ref{eq:secstep21}), and (\ref{eq:secstep26}).
\end{proof}
The results for the sectional threshold obtained from the above theorem as well as the best currently known ones from \cite{DonohoPol,DonohoUnsigned}
are presented on Figure \ref{fig:sec}. As can be seen, the threshold results obtained from the previous analysis slightly improve on those from
\cite{DonohoPol,DonohoUnsigned}.

%%%%%%%%%%%%%%%%%%%%%%%%%%%%%%%%%%%%%%%%%%%%%%%%%%%%%%%%%%%%%%%%%%%%%%%%%%%%%%%%
\section{Probabilistic analysis of the null-space characterizations -- signed $\x$}
\label{sec:signed}
%%%%%%%%%%%%%%%%%%%%%%%%%%%%%%%%%%%%%%%%%%%%%%%%%%%%%%%%%%%%%%%%%%%%%%%%%%%%%%%%

In this section we consider recovery of vectors $\x$ with elements known to have certain sign pattern. Without loss of generality we assume that it is known that $\x_i\geq 0,1\leq i\leq n$. We also again assume that $\x$ is $k$-sparse, i.e. we assume that $\x$ has no more than $k$ nonzero elements. To solve (\ref{eq:system}) for such an $\x$ instead of (\ref{eq:l1}) we consider the following optimization problem
\begin{eqnarray}
\mbox{min} & & \|\x\|_{1}\nonumber \\
\mbox{subject to} & & A\x=\y\nonumber \\
& & \x_i\geq 0. \label{eq:l1signed}
\end{eqnarray}
The following theorem from e.g. \cite{StojnicICASSP09} characterizes the equivalence of (\ref{eq:system}) and (\ref{eq:l1signed}).
\begin{theorem}(Null-space characterization; Non-negative $\x$)
Assume that an $m\times n$ measurement matrix $A$ is given. Let $\x$
be a $k$-sparse vector whose non-zero components are known to be positive.
Further, assume that $\y=A\x$ and that $\w$ is an $n\times 1$
vector. Let $K$ be any subset of $\{1,2,\dots,n\}$ such that $|K|=k$
and let $K_i$ denote the $i$-th element of $K$. Further, let
$\bar{K}=\{1,2,\dots,n\}\setminus K$. Then (\ref{eq:l1signed}) will
produce the solution of (\ref{eq:system}) if
\begin{eqnarray}
&\forall K \quad \mbox{and} \quad (\forall \w\in \textbf{R}^n |
A\w=0,\w_{\bar{K}_i}\geq 0,1\leq i\leq n-k),&\quad\nonumber \\
&-\sum_{i=1}^{k}\w_{K_i}<\sum_{i=1}^{n-k}\w_{\bar{K}_i}.&
\label{eq:thmeqnon}
\end{eqnarray}
\label{thm:thmgennon}
\end{theorem}
In the rest of this section we will probabilistically analyze validity of (\ref{eq:thmeqnon}) (or to be more precise, its a slight modification).
In the first following subsection we will show how one can obtain the values of the weak threshold $\betawnon$ for the entire range $0\leq \alpha\leq 1$ based on such a probabilistic analysis.
%In the second subsection we obtain the values of the strong threshold.

%%%%%%%%%%%%%%%%%%%%%%%%%%%%%%%%%%%%%%%%%%%%%%%%%%%%%%%%%%%
\subsection{Weak threshold} \label{sec:weaksigned}
%%%%%%%%%%%%%%%%%%%%%%%%%%%%%%%%%%%%%%%%%%%%%%%%%%%%%%%%%%%

In this subsection we determine the weak threshold $\betawnon$. Before proceeding further we quickly recall on the definition of the weak threshold. The definition of the weak threshold was already introduced in Section \ref{sec:weakgen} when recovery of general signals (vectors) $\x$ was considered. Here, we slightly modify it so that it fits the scenario of a priori known sign patterns of elements of $\x$. Namely, for a given $\alpha$, $\betawnon$ is the maximum value of $\beta$ such that the solutions of (\ref{eq:system}) and (\ref{eq:l1signed}) coincide for any given $\beta n$-sparse $\x$ with a fixed location of nonzero components and a priori known to be comprised of non-negative elements. Since the analysis will clearly be irrelevant with respect to what particular location of nonzero elements is chosen, we can for the simplicity of the exposition and without loss of generality assume that the components $\x_{1},\x_{2},\dots,\x_{n-k}$ of $\x$ are equal to zero and the components $\x_{n-k+1},\x_{n-k+2},\dots,\x_n$ of $\x$ are greater than or equal to zero. Under this assumption we have the following corollary of Theorem \ref{thm:thmgennon}.
\begin{corollary}(Nonzero part of $\x$ has fixed a location; The signs of elements of $\x$ a priori known)
Assume that an $m\times n$ measurement matrix $A$ is given. Let $\x$
be a $k$-sparse vector whose nonzero components are known to be positive. Also let $\x_1=\x_2=\dots=\x_{n-k}=0.$
Further, assume that $\y=A\x$ and that $\w$ is
an $n\times 1$ vector. Then (\ref{eq:l1signed}) will
produce the solution of (\ref{eq:system}) if
\begin{equation}
(\forall \w\in \textbf{R}^n | A\w=0, \w_{i}\geq 0,1\leq i\leq n-k) \quad  -\sum_{i=n-k+1}^n \w_i<\sum_{i=1}^{n-k}\w_{i}.
\label{eq:thmeqgennonweak}
\end{equation}
\label{thm:thmgennonweak}
\end{corollary}
Following the procedure of Subsection \ref{sec:weakgen} we set
$\Swnon$
\begin{equation}
\Swnon=\{\w\in S^{n-1}| \quad \w_{i}\geq 0,1\leq i\leq n-k \quad \mbox{and}\quad -\sum_{i=n-k+1}^n \w_i<\sum_{i=1}^{n-k}\w_{i}\}\label{eq:defSwnon}
\end{equation}
and
\begin{equation}
w(\Swnon)=E\sup_{\w\in \Swnon} (\h^T\w) \label{eq:widthdefSwnon}
\end{equation}
where as earlier $\h$ is a random column vector in $R^{n}$ with i.i.d. ${\cal
N}(0,1)$ components and $S^{n-1}$ is the unit $n$-dimensional sphere. As in Subsection \ref{sec:stronggen} our goal will be to compute an upper bound on $w(\Swnon)$ and then equal that upper bound to $\left ( \sqrt{m}-\frac{1}{4\sqrt{m}}\right )$. To simplify the exposition we again set $w(\h,\Swnon)=\max_{\w\in \Swnon} (\h^T\w)$. We will proceed again as earlier and in Subsection \ref{sec:weakubnon} we will determine an upper bound $\Bweaknon$ on $w(\h,\Swnon)$. In Subsection \ref{sec:weakexpnon} we will compute an upper bound on $E (\Bweaknon)$. That quantity will be an upper bound on $w(\Swnon)$ since according to the following $E(\Bweaknon)$ is an upper bound on $w(\Swnon)$
\begin{equation}
w(\Swnon)=Ew(\h,\Swnon)=E(\max_{\w\in \Swnon} (\h^T\w))\leq E(\Bweaknon).\label{eq:weakgoalnon}
\end{equation}

%%%%%%%%%%%%%%%%%%%%%%%%%%%%%%%%%%%%%%%%%%%%%%%%%%%%%%%%%%%%%%
\subsubsection{Upper-bounding $w(\h,\Swnon)$}
\label{sec:weakubnon}
%%%%%%%%%%%%%%%%%%%%%%%%%%%%%%%%%%%%%%%%%%%%%%%%%%%%%%%%%%%%%%%

In a fashion analogous to (\ref{eq:workws0}) we can write
\begin{equation}
w(\h,\Swnon)=\max_{\w\in \Swnon} (\h^T\w)=\max_{\w\in \Swnon} (\sum_{i=1}^{n-k} \h_i\w_i+\sum_{i=n-k+1}^n\h_i\w_i).\label{eq:workww0non}
\end{equation}
Let again $\h_{1:(n-k)}=(\h_1,\h_2,\dots,\h_{n-k})^T$. Further, let
$\h_{(i)}^{(n-k)}$ be the $i$-th smallest of the elements of $\h_{1:(n-k)}$. Set
\begin{equation}
\hwnon=(\h_{(1)}^{(n-k)},\h_{(2)}^{(n-k)},\dots,\h_{(n-k)}^{(n-k)},-\h_{n-k+1},-\h_{n-k+2},\dots,-\h_{n})^T.\label{eq:defhwnon}
\end{equation}
Then one can simplify (\ref{eq:workww0non}) in the following way
\begin{eqnarray}
w(\h,\Sw) = \max_{\y\in R^n} & &  (\hwnon)^T\y_i \nonumber \\
\mbox{subject to} &  & \y_i\geq 0, 0\leq i\leq (n-k)\nonumber \\
& & \sum_{i=n-k+1}^n\y_i\geq \sum_{i=1}^{n-k} \y_i \nonumber \\
& & \sum_{i=1}^n\y_i^2\leq 1.\label{eq:workww1}
\end{eqnarray}
One can then proceed in the similar fashion as in Subsection \ref{sec:strongub} and compute an upper bound based on duality. The only differences are that we now have $\hwnon$ instead of $\tilh$ and positive components of $\y$ are only those with indexes less than or equal to $(n-k)$. After repeating one more time literally every step of the derivation from Subsection \ref{sec:strongub} one obtains the following analogue to the equation (\ref{eq:strfinalub})
\begin{equation}
w(\h,\Swnon)\leq\sqrt{\sum_{i=1}^n|\hwnon_i|^2-\sum_{i=1}^{c}|\hwnon_i|^2-\frac{(((\hwnon)^T\z)-\sum_{i=1}^{c}\hwnon_i)^2}{n-c}}
=\sqrt{\sum_{i=c+1}^n|\hwnon_i|^2-\frac{(((\hwnon)
^T\z)-\sum_{i=1}^{c}\hwnon_i)^2}{n-c
}}\label{eq:weakfinalubnon}
\end{equation}
where $\hwnon_i$ is the $i$-th element of $\hwnon$ and $c\leq (n-k)$ is such that $(((\hwnon)
^T\z)-\sum_{i=1}^{c}\hwnon_i)\geq 0$. Clearly, as long as $((\hwnon)^T\z)\geq 0$ there will be a $c$ (it is possible that $c=0$) such that quantity on the most right hand side of
(\ref{eq:weakfinalubnon}) is an upper bound on $w(\h,\Swnon)$.

Using (\ref{eq:weakfinalubnon}) we then establish the following analogue to Lemma \ref{lm:Bstr}.
\begin{lemma}
Let $\h\in R^n$ be a vector with i.i.d. zero-mean unit variance gaussian components. Further let $\hwnon$ be as defined in (\ref{eq:defhwnon}) and $w(\h,\Swnon)=\max_{\w\in \Swnon} (\h^T\w)$ where $\Swnon$ is as defined in (\ref{eq:defSwnon}). Let $\z\in R^n$ be a column vector such that $\z_i=1,1\leq i\leq (n-k)$ and $\z_i=-1,n-k+1\leq i\leq n$. Then
\begin{equation}
w(\h,\Swnon)\leq \Bweaknon \label{eq:weaklemmaubnon}
\end{equation}
where
\begin{equation}
\Bweaknon=
\begin{cases}
\sqrt{\sum_{i=1}^n|\hwnon_{i}|^2} & \mbox{if} \quad \zeta_{w}^+(\h,\cweaknon)\leq 0 \\
\sqrt{\sum_{i=\cweaknon+1}^n|\hwnon_{i}|^2-\frac{(((\hwnon)^T\z)-\sum_{i=1}^{\cweaknon}\hwnon_{i})^2}{n-\cweaknon}} & \mbox{if} \quad \zeta_{w}^+(\h,\cweaknon)>0
\end{cases},
\end{equation}
$\zeta_{w}^+(\h,c)=\frac{((\hwnon)^T\z)-\sum_{i=1}^c\hwnon_{i}}{n-c}-\hwnon_{c}$
and $\cweaknon=\deltawnon n$ is a $c\leq n-k$ such that
\begin{equation}
\frac{(1-\epsilon)E(((\hwnon)^T\z)-\sum_{i=1}^c\hwnon_{i})}{n-c}-F_c^{-1}\left (\frac{(1+\epsilon)c}{n(1-\betawnon)}\right )= 0.
\end{equation}
$F_c^{-1}(\cdot)$ is the inverse cdf of zero-mean, unit variance Gaussian random variable. $\epsilon>0$ is an arbitrarily small constant independent of $n$.\label{lm:Bweaknon}
\end{lemma}
\begin{proof}
Follows directly from the derivation before Lemma \ref{lm:Bstr} replacing $\tilh$ by $\hwnon$.
\end{proof}

%%%%%%%%%%%%%%%%%%%%%%%%%%%%%%%%%%%%%%%%%%%%%
\subsubsection{Computing  an upper bound on $E(\Bweaknon)$} \label{sec:weakexpnon}
%%%%%%%%%%%%%%%%%%%%%%%%%%%%%%%%%%%%%%%%%%%%%

Following step-by-step the derivation of Lemma \ref{lm:EBstr} we can establish the following analogue to it.
\begin{lemma}
Assume the setup of Lemma \ref{lm:Bweaknon}. Let further $\psi_w^+=\frac{E((\hwnon)^T\z)-\sum_{i=1}^{\cweaknon}\hwnon_{i})}{n}$.Then
\begin{equation}
\hspace{-.7in}E(\Bweaknon)\leq \sqrt{n}\left
(\exp\left \{-\frac{n\epsilon^2\deltawnon}{2(1+\epsilon)} \right \}+\exp \left \{  -\frac{\epsilon^2 (\psi_w^+)^2 n}{2} \right \}\right )
+\sqrt{E\sum_{i=\cweaknon+1}^n|\hwnon_{i}|^2-\frac{(E((\hwnon)^T\z)-E\sum_{i=1}^{\cweaknon}\hwnon_{i})^2}{n-\cweaknon}}.\label{eq:weakexp9non}
\end{equation}\label{lm:EBweaknon}
\end{lemma}
\begin{proof}
Follows directly from the derivation before Lemma \ref{lm:EBstr}.
\end{proof}

As earlier, following (\ref{eq:stralpha}), if $n$ is large, for a fixed $\alpha$ one can determine $\betawnon$ as a maximum $\beta$ such that
\begin{equation}
\alpha\geq \frac{E\sum_{i=\cweaknon+1}^n|\hwnon_{i}|^2}{n}-\frac{(E((\hwnon)^T\z)-E\sum_{i=1}^{\cweaknon}\hwnon_{i})^2}{n(n-\cweaknon)}.\label{eq:weakalphanon}
\end{equation}
In the rest of this subsection we show how the left hand side of (\ref{eq:weakalphanon}) can be computed for a randomly chosen fixed $\betawnon$. We will again repeat the two crucial steps:
\begin{enumerate}
\item We first determine $\cweaknon$
\item We then compute $\frac{E\sum_{i=\cweaknon+1}^n|\hwnon_{i}|^2}{n}-\frac{(E((\hwnon)^T\z)-E\sum_{i=1}^{\cweaknon}\hwnon_{i})^2}{n(n-\cweaknon)}$ with $\cweaknon$ found
in step $1$.
\end{enumerate}

\underline{\emph{Step 1:}}

From Lemma \ref{lm:Bweaknon} we have $\cweaknon=\deltawnon n$ is a $c$ such that
\begin{eqnarray}
& & \frac{(1-\epsilon)E((\sum_{i=1}^{n-\betawnon n}\hwnon_{i}-\sum_{i=n-\betawnon n+1}^n\hwnon_{i})-\sum_{i=1}^{\deltawnon n}\hwnon_{i})}{n-c}-F_c^{-1}\left (\frac{(1+\epsilon)c}{n(1-\betawnon)}\right )= 0\nonumber     \\
& \Leftrightarrow & \frac{(1-\epsilon)(E\sum_{i=1}^{n-\betawnon n}\hwnon_i+E\sum_{i=n-\betawnon n+1}^n\h_{i}-E\sum_{i=1}^{\deltawnon n}\hwnon_i)}{n-c}-F_c^{-1}\left (\frac{(1+\epsilon)c}{n(1-\betawnon)}\right )= 0\nonumber \\\label{eq:weakdelta1non}
\end{eqnarray}
where we recall that now $\hwnon_i,1\leq i\leq (n-\betawnon n)$, is the $i$-th smallest element (not magnitude) of vector $\h_{1:(n-\betawnon n)}$. Also, we easily have
$E\h_i=0,n-\betawnon n+1\leq i\leq n$, and then from (\ref{eq:weakdelta1non})
\begin{eqnarray}
& & \frac{(1-\epsilon)E((\sum_{i=1}^{n-\betawnon n}\hwnon_{i}-\sum_{i=n-\betawnon n+1}^n\hwnon_{i})-\sum_{i=1}^{\deltawnon n}\hwnon_{i})}{n-c}-F_c^{-1}\left (\frac{(1+\epsilon)c}{n(1-\betawnon)}\right )= 0\nonumber \\
& \Leftrightarrow & \frac{(1-\epsilon)E\sum_{i=\deltawnon n+1}^{n-\betawnon n}\hwnon_i}{n(1-\deltawnon)}-F_c^{-1}\left (\frac{(1+\epsilon)\deltawnon n}{n(1-\betawnon)}\right )= 0.\label{eq:weakdelta2non}
\end{eqnarray}
Set $\thetawnon=1-\deltawnon$. Following \cite{Stigler73,BarvSam} and (\ref{eq:strstep22}) we obtain
\begin{equation}
\lim_{n\rightarrow\infty}\frac{E\sum_{i=(1-\thetawnon) n+1}^{(1-\betawnon)n}\hwnon_i}{n(1-\betawnon)} = \int_{F_c^{-1}(\frac{1-\thetawnon}{1-\betawnon})}^{\infty}tdF_c(t).\label{eq:weakdelta3non}
\end{equation}
We first easily compute $F_c^{-1}(\frac{1-\thetawnon}{1-\betawnon})$ in the following way
\begin{eqnarray}
& & \frac{1}{\sqrt{2\pi}}\int_{\-\infty}^{F_c^{-1}(\frac{1-\thetawnon}{1-\betawnon})}e^{-\frac{t^2}{2}}dt=\frac{1-\thetawnon}{1-\betawnon}\nonumber \\
& \Longrightarrow & F_c^{-1}(\frac{1-\thetawnon}{1-\betawnon})=\sqrt{2}\erfinv\left (2\frac{1-\thetawnon}{1-\betawnon}-1\right ).\label{eq:weakdelta4non}
\end{eqnarray}
In a similar fashion one then has
\begin{equation}
F_c^{-1}\left (\frac{(1+\epsilon)\deltawnon n}{n(1-\betawnon)}\right )=\sqrt{2}\erfinv\left (2\frac{(1+\epsilon)(1-\thetawnon)}{1-\betawnon}-1\right ).
\label{eq:weakdelta5non}
\end{equation}
Using (\ref{eq:weakdelta4non}) we further find
\begin{equation}
\int_{F_c^{-1}(\frac{1-\thetawnon}{1-\betawnon})}^{\infty}tdF_c(t)=\sqrt{\frac{1}{2\pi}}
\int_{F_c^{-1}(\frac{1-\thetawnon}{1-\betawnon})}^{\infty}te^{-\frac{t^2}{2}}dt=\sqrt{\frac{1}{2\pi}}
e^{-\erfinv\left (2\frac{1-\thetawnon}{1-\betawnon}-1\right )^2}.\label{eq:weakdelta6non}
\end{equation}
Combining (\ref{eq:weakdelta1non}), (\ref{eq:weakdelta2non}), (\ref{eq:weakdelta3non}), (\ref{eq:weakdelta5non}), and (\ref{eq:weakdelta6non}) we obtain the following equation for finding $\thetawnon$
\begin{equation}
(1-\epsilon)(1-\betawnon)\frac{\sqrt{\frac{1}{2\pi}}
e^{-\erfinv\left (2\frac{1-\thetawnon}{1-\betawnon}-1\right )^2}}{\thetawnon}-\sqrt{2}\erfinv\left (2\frac{(1+\epsilon)(1-\thetawnon)}{1-\betawnon}-1\right )=0.
\label{eq:weakthetanon}
\end{equation}
Let $\hthetawnon$ be the solution of (\ref{eq:weakthetanon}). Then $\deltawnon=1-\hthetawnon$ and $\cweaknon=\deltawnon n=(1-\hthetawnon)n$. This concludes step $1$.

\underline{\emph{Step $2$:}}

In this step we compute
$\frac{E\sum_{i=\cweaknon+1}^n|\hwnon_{i}|^2}{n}-\frac{(E((\hwnon)^T\z)-E\sum_{i=1}^{\cweaknon}\hwnon_{i})^2}{n(n-\cweaknon)}$ with $\cweaknon=(1-\hthetawnon)n$.
Using results from step $1$ we easily find
\begin{equation}
\lim_{n\rightarrow\infty}\frac{(E((\hwnon)^T\z)-E\sum_{i=1}^{\cweaknon}\hwnon_{i})^2}{n(n-\cweaknon)}=
\frac{\left ((1-\betawnon)\sqrt{\frac{1}{2\pi}}e^{-(\erfinv(2\frac{1-\hthetawnon}{1-\betawnon}-1))^2}\right )^2}{\hthetawnon}.\label{eq:weakstep21non}
\end{equation}
Effectively, what is left to compute is $\frac{E\sum_{i=\cweaknon+1}^n|\hwnon_{i}|^2}{n}$. First we note that
\begin{equation}
\frac{E\sum_{i=\cweaknon+1}^n|\hwnon_{i}|^2}{n}=\frac{E\sum_{i=(1-\hthetawnon)n+1}^{(1-\betawnon)n}|\hwnon_i|^2+E\sum_{i=(1-\betawnon)n+1}^{n}\h_{i}^2}{n}
=\frac{E\sum_{i=(1-\hthetawnon)n+1}^{(1-\betawnon)n}|\hwnon_i|^2}{n}
+\betawnon.\label{eq:weakstep22non}
\end{equation}
Using an approach similar to the one from step $2$ of Subsection \ref{sec:weakexp} and following \cite{Stigler73,BarvSam} we have
\begin{equation}
\lim_{n\rightarrow\infty}\frac{E\sum_{i=(1-\hthetawnon)n+1}^{(1-\betawnon)n}|\hwnon_i|^2}{n(1-\betawnon)}=
\int_{F_d^{-1}(\frac{1-\hthetawnon}{1-\betawnon})}^{\infty}tdF_d(t)\label{eq:weakstep23non}
\end{equation}
where $F_d^{-1}$ is the inverse cdf of random variable $\mbox{sign} (X)|X|^2$ and $X$ is zero-mean unit variance Gaussian random variable.
Straightforward calculations produce
\begin{eqnarray}
F_d^{-1}(\frac{1-\hthetawnon}{1-\betawnon})=2(\erfinv(2\frac{1-\hthetawnon}{1-\betawnon}-1))^2\label{eq:weakstep24non}
\end{eqnarray}
and
\begin{equation}
\int_{F_d^{-1}(\frac{1-\hthetawnon}{1-\betawnon})}^{\infty}tdF_d(t)=\frac{1}{2}\frac{1}{\sqrt{2\pi}}\left (\sqrt{2\pi}+2\frac{\sqrt{F_d^{-1}(\frac{1-\hthetawnon}{1-\betawnon})}}{\exp\left \{ \frac{F_d^{-1}(\frac{1-\hthetawnon}{1-\betawnon})}{2}\right \}}-\sqrt{2\pi}\left (2\frac{1-\hthetawnon}{1-\betawnon}-1\right )\right ).\label{eq:weakstep25non}
\end{equation}
Combining (\ref{eq:weakstep22non}), (\ref{eq:weakstep23non}), (\ref{eq:weakstep24non}), and (\ref{eq:weakstep25non}) we obtain
\begin{eqnarray}
\hspace{-.6in}\lim_{n\rightarrow\infty}\frac{E\sum_{i=(1-\hthetawnon) n+1}^{n}|\hwnon_{i}|^2}{n}  =  \frac{1-\betawnon}{2\sqrt{2\pi}}\left (\sqrt{2\pi}+2\frac{\sqrt{2(\erfinv(2\frac{1-\hthetawnon}{1-\betawnon}-1))^2}}{e^{(\erfinv(2\frac{1-\hthetawnon}{1-\betawnon}-1))^2}}-\sqrt{2\pi}
(2\frac{1-\hthetawnon}{1-\betawnon}-1)\right )+\betawnon.\nonumber \\\label{eq:weakstep26non}
\end{eqnarray}

We summarize the results from this section in the following theorem.

\begin{theorem}(Weak threshold, a priori known signs of $\x$)
Let $A$ be an $m\times n$ measurement matrix in (\ref{eq:system})
with the null-space uniformly distributed in the Grassmanian. Let
the unknown $\x$ in (\ref{eq:system}) be $k$-sparse. Let it be known that the nonzero components of $\x$ are positive. Further, let the locations of nonzero elements of $\x$ be arbitrarily chosen but fixed.
Let $k,m,n$ be large
and let $\alpha=\frac{m}{n}$ and $\betawnon=\frac{k}{n}$ be constants
independent of $m$ and $n$. Let $\erfinv$ be the inverse of the standard error function associated with zero-mean unit variance Gaussian random variable.  Further,
let $\epsilon>0$ be an arbitrarily small constant and $\hthetawnon$, ($\betawnon\leq \hthetawnon\leq 1$), be the solution of
\begin{equation}
(1-\epsilon)(1-\betawnon)\frac{\sqrt{\frac{1}{2\pi}}e^{-(\erfinv(2\frac{1-\thetawnon}{1-\betawnon}-1))^2}}{\thetawnon}-\sqrt{2}\erfinv ((2\frac{(1+\epsilon)(1-\thetawnon)}{1-\betawnon}-1))=0.\label{eq:thmweakthetanon}
\end{equation}
If $\alpha$ and $\betawnon$ further satisfy
\begin{equation}
\hspace{-.7in}\alpha>\frac{1-\betawnon}{2\sqrt{2\pi}}\left (\sqrt{2\pi}+2\frac{\sqrt{2(\erfinv(2\frac{1-\hthetawnon}{1-\betawnon}-1))^2}}{e^{(\erfinv(2\frac{1-\hthetawnon}{1-\betawnon}-1))^2}}-\sqrt{2\pi}
\left (2\frac{1-\hthetawnon}{1-\betawnon}-1\right )\right )+\betawnon
-\frac{\left ((1-\betawnon)\sqrt{\frac{1}{2\pi}}e^{-(\erfinv(2\frac{1-\hthetawnon}{1-\betawnon}-1))^2}\right )^2}{\hthetawnon}\label{eq:thmweakalphanon}
\end{equation}
then the solutions of (\ref{eq:system}) and (\ref{eq:l1signed}) coincide with overwhelming
probability.\label{thm:thmweakthrnon}
\end{theorem}
\begin{proof}
Follows from the previous discussion combining (\ref{eq:thmesh}), (\ref{eq:weakgoalnon}), (\ref{eq:weaklemmaubnon}), (\ref{eq:weakexp9non}), (\ref{eq:weakalphanon}), (\ref{eq:weakthetanon}), (\ref{eq:weakstep21non}), and (\ref{eq:weakstep26non}).
\end{proof}

\begin{figure}[htb]
%%%%%\begin{minipage}[b]{1.0\linewidth}
\centering
\centerline{\epsfig{figure=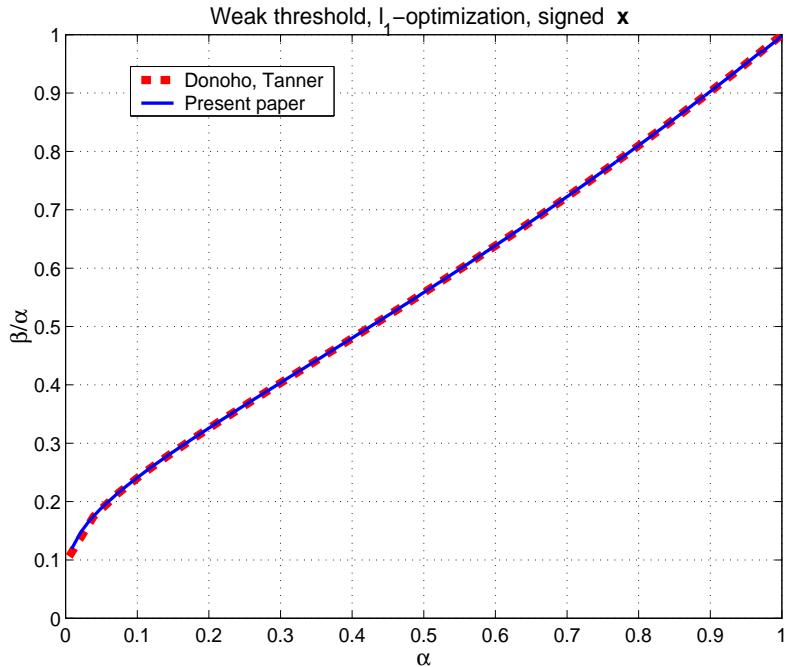,width=10.5cm,height=9cm}}
%%%%%%\end{minipage}
\vspace{-0.2in} \caption{\emph{Weak} threshold, $\ell_1$-optimization; signed $\x$}
\label{fig:weaksigned}
\end{figure}

The results for the weak thresholds obtained from the above theorem in the case of a priori known signs of components of $\x$ as well as the best currently known ones from \cite{DT,DonohoSigned}
are presented on Figure \ref{fig:weaksigned}. As can be seen, the threshold results obtained from the previous analysis match those from \cite{DT,DonohoSigned}.

%%%%%%%%%%%%%%%%%%%%%%%%%%%%%%%%%%%%%%%%%%%%%%%%%%%%%%%%%%%%%%%%%%%%%%%%%%%%%%%%
\section{Discussion}
\label{sec:discuss}
%%%%%%%%%%%%%%%%%%%%%%%%%%%%%%%%%%%%%%%%%%%%%%%%%%%%%%%%%%%%%%%%%%%%%%%%%%%%%%%%

In this paper we considered recovery of sparse signals from a reduced number of linear measurements.
We provided a theoretical performance analysis of a classical polynomial
$\ell_1$-optimization algorithm. Under the assumption that the measurement matrix $A$ has a basis of the null-space
distributed uniformly in the
Grassmanian, we derived lower bounds on the values of the recoverable strong, weak, and sectional
thresholds in the so-called linear regime, i.e. in the regime when
the recoverable sparsity is proportional to the length of the
unknown vector. Obtained threshold results are comparable to the best currently known ones.

The main subject of this paper was recovery of the so-called ideally sparse signals.
It is not that difficult to see that the presented analysis
framework is very general and can be extended to
include computations of threshold values for recovery of
approximately sparse signals as well as those with noisy
measurements. We should also mention that in this paper we were
mostly concerned with the success of $\ell_1$-optimization. However,
the results similar to those presented in this paper
can be obtained in the case of $\ell_q$-optimization ($0<q<1$) as well.
All these generalizations will be part of a future work.

Carefully following our exposition one can note that the strong threshold results in case of signed vectors $\x$ are missing. We should mention that the procedure presented in this paper can be repeated for that case as well. However, due to a somewhat complicated structure of the set $S$ the analysis of that case becomes a bit more tedious and certainly loses on elegance. Nevertheless, we conducted the analysis. However, since the final results that we obtained trail those of \cite{DT,DonohoSigned} (except in a very narrow range around $\alpha\longrightarrow 1$) we decided not to include them in this paper.

On the technical side we should mention that our analysis made critical use of an excellent work \cite{Gordon88}.
On the other hand \cite{Gordon88} massively relied on phenomenal results
\cite{Pisier86,CIS76} related to the estimates of the normal tail
distributions of Lipshitz functions. In a very recent work related
to the matrix-rank optimization the authors in \cite{RXH08}
successfully applied results of \cite{Pisier86,CIS76} directly
without relying on the main results from \cite{Gordon88}. It
will certainly be interesting to see what performance guarantees the
direct application of the results of \cite{Pisier86,CIS76} would
produce in the problems considered in this paper.

At the end we should finally mention a potential universal value of
the results presented here. In this paper we were mostly concerned
with the compressed sensing signal processing applications. However,
the results presented here may be of independent mathematical
interest as well. First, clearly our analysis (as almost any other
analysis related to compressed sensing) has immediate impact on
important mathematical problem of solving under-determined systems
of linear equations. Second, following the derivations of
\cite{DonohoPol,DT,DTbern} it is not that difficult to see that our
results can be directly applied to determine the neighborliness thresholds of
projected cross-polytope, regular simplex, and positive orthant as well.

%\newpage1
%\setcounter{page}{1}
\begin{singlespace}
\bibliographystyle{plain}
\bibliography{CSetamRefs}
\end{singlespace}

\end{document}